\documentclass[pdflatex,sn-nature]{sn-jnl}%

\usepackage{graphicx}%
\usepackage{multirow}%
\usepackage{amsmath,amssymb,amsfonts}%
\usepackage{amsthm}%
\usepackage{mathrsfs}%
\usepackage[title]{appendix}%
\usepackage{xcolor}%
\usepackage{textcomp}%
\usepackage{manyfoot}%
\usepackage{booktabs}%
\usepackage{tabularx}%
\usepackage{algorithm}%
\usepackage{algorithmicx}%
\usepackage{algpseudocode}%
\usepackage{listings}%

\theoremstyle{thmstyleone}%

\theoremstyle{thmstyletwo}%

\theoremstyle{thmstylethree}%

\begin{document}

\title[How Did Data Sharing Evolve?]{\hspace{1.6cm} 
How did data sharing evolve\\ in the decade since the FAIR Principles?}

\author*[1]{\fnm{Sheeba} \sur{Samuel}}\email{sheeba.samuel@informatik.tu-chemnitz.de}

\author*[2,3]{\fnm{Daniel} \sur{Mietchen}}\email{daniel.mietchen@fiz-karlsruhe.de}

\affil*[1]{\orgdiv{Distributed and Self-organizing Systems}, \orgname{Chemnitz University of Technology}, \orgaddress{\street{Straße der Nationen}, \city{Chemnitz}, \postcode{09111}, \state{Sachsen}, \country{Germany}}}

\affil[2]{\orgdiv{Department of Mathematics}, \orgname{FIZ Karlsruhe - Leibniz Institute for Information Infrastructure}, \orgaddress{\street{Franklinstr. 11}, \postcode{10587}, \state{Berlin}, \country{Germany}}}

\affil[3]{\orgname{Institute for Globally Distributed Open Research and Education (IGDORE)}, \orgaddress{\street{Medicinaregatan 8A, Plan 4}, \city{G{\"o}teborg}, \postcode{413 90}, \country{Sweden}}}

\abstract{

A decade after the publication of the FAIR Guiding Principles, what have we learned about data sharing, and what do data availability statements (DAS) and other materials that have accrued over that period tell us about how data-sharing practices, attitudes, and infrastructures have evolved?
Here, we present a large-scale longitudinal analysis of over 6.8~million biomedical articles from the PubMed Central Open Access Subset, spanning the period from 2011 till 2025. 
Using an automated 
pipeline, we extracted and classified DAS, repository links, accession numbers, and supplementary materials, assigning each article a data-sharing score and independently FAIR-assessing a curated subset of cited dataset URLs. 
We find that, in relative terms, explicit DAS adoption rose from approximately 2\% of articles in 2011 to 68\% in 2025,
while statements of the ``available on request'' type 
grew in parallel from near zero to over 19\%.
The corpus's growth amplifies both patterns: in absolute terms, 
articles with a data availability statement containing an explicit mention of a data repository rose from 132 in 2011 to 94,697 in 2025 (a $\sim$720-fold increase), while ``available on request'' articles rose from 77 to 173,474 (i.e.\ $>$2,250-fold) over the same 2011--2025 span.
Structural biology and genomics consistently achieved the highest data-sharing scores, while clinical research and public health ranked lowest.
Gold Open Access journals showed a small overall advantage over subscription journals, 
while Diamond Open Access journals -- which charge neither readers nor authors -- clearly outperformed both.
Another 
observation is that internationally collaborative articles scored higher than domestic ones.  Internationally collaborative articles scored higher than domestic ones. A gender gap emerged as well: female first-authored articles' scores declined more than male first-authored ones'. 
The analysis further included 
additional dimensions, from compliance with funder and journal mandates or with JATS XML tagging standards to computational reproducibility signals and the co-occurrence of DAS with ethics statements.
For instance, ethics statements, present in 17.9\% of articles, are associated with a higher DAS rate but a lower mean data-sharing score, consistent with human-subjects research more often falling into restrictive, privacy-centric sharing categories.
Taken together, a decade on, the gap between declaring data availability and actually sharing machine-readable, reusable data persists, albeit with variation across fields, infrastructures and demographics.

}


\maketitle

\section{Introduction}
\label{sec:introduction}
The FAIR Guiding Principles (Findability, Accessibility, Interoperability, and Reusability) for the sharing of research data and associated metadata \cite{wilkinson2016fair} were introduced as a stakeholder-neutral framework for data management and stewardship, published in March 2016 after development within the FORCE11 community. Rather than a technical standard or certification, the principles were deliberately formulated as a minimal, high-level set of guideposts, applicable regardless of discipline, that data and metadata should satisfy to be genuinely machine-actionable and reusable by both humans and computational workflows.

In 2018, two years after the principles' publication, a design framework and exemplar maturity indicators followed \cite{wilkinson2018a}, giving rise to a family of assessment approaches. Early FAIR self-assessment relied on manual questionnaires and expert-reviewed checklists; these did not scale to corpus-wide application and suffered from inter-rater inconsistency, motivating a shift toward automated, rule-based evaluation such as the currently popular F-UJI \cite{devaraju2021an}.

In parallel and building
on FAIR-inspired prior work on data citation \cite{mietchen2015adapting},
the JATS4R community initiative \cite{beck2015improving}
published prescriptive JATS XML markup recommendations for data-availability \cite{jats4r2018das} and ethics-statement \cite{niso2020ethics} sections, aiming to close the gap between human-readable and machine-actionable statements~-- a distinction that this study tracks throughout as ``compliance'' versus mere ``presence'' (cf.\ Sections \ref{sec:automatedassessment} and \ref{sec:JATS4R}). 
Of note, both F-UJI and JATS4R focus on assessing compliance at a given moment rather than on changes over time. Corpus-scale literature-mining proposals aimed specifically at structured, longitudinally trackable statement content, such as \cite{Hegde2022Mining}'s 
approach to mining ethics statements, and recent work tracking the reproducibility of released research software over time \cite{Coakley2026shift}, point toward 
addressing such gaps
outside the FAIR-assessment tools themselves. 

Before the FAIR Principles existed, evidence was already accumulating that data-sharing \emph{statements} do not reliably translate into data-sharing \emph{practice}. Vines et al. \cite{vines2014the} showed that the odds of a dataset still being available fall by roughly 17\% per year after publication, and Federer et al. \cite{federer2018data} found that only a minority of PLOS ONE 
Data Availability Statements pointed to a usable, non-supplementary data location even under one of the field's strictest mandates. More recently, the ``available upon request'' pattern was tested
directly by emailing corresponding authors, which led to fewer than two in five \cite{tedersoo2021data} and fewer than one in five \cite{hussey2025data} requests, respectively, being honoured.
Likewise, \cite{dunic2026do} question whether the mere presence of a DAS increases actual data sharing; both approaches offer direct behavioural evidence for the anti-pattern this study tracks at corpus scale in Section \ref{sec:fromsilence} and Section \ref{sec:funder}. 
At the corpus level, Wallach et al. \cite{wallach2018reproducible} audited a large sample of the biomedical literature (2015--2017) against multiple transparency indicators and, like this study, found improvement concentrated in a subset of indicators rather than across the board. Code sharing alone does not guarantee computational reproducibility, a point this study's own prior work on Jupyter notebooks from biomedical publications makes directly -- both in auditing their actual re-execution success \cite{samuel2024computational} and in a separate reproducibility-assessment framework \cite{samuel2026reproscore} -- and that Section~\ref{sec:limits} below confirms at corpus scale: the mean reproducibility-signal uplift associated with code sharing is large, but only 2.1\% of the corpus carries any reproducibility signal at all. 
Colavizza et al. \cite{colavizza2020citation} showed that articles linking to a data repository accumulate a measurable citation advantage, one candidate mechanism for the ``named repository'' DAS category's growth documented in Section \ref{sec:fromsilence} and Section \ref{sec:repositorychoice}.

The disciplinary and demographic unevenness of open-science uptake is likewise established outside the data-sharing literature specifically. Huang et al. \cite{huang2020historical} documented a persistent, if narrowing, gender gap in scientific productivity and career length across countries and fields, a structural backdrop against which the first-authorship data-sharing-score gap reported in Section \ref{sec:persistentgender} should be read, since neither study can separate a genuine practice gap from compounding factors such as field, seniority, and institutional access. On the publishing-model side, Bosman et al. \cite{bosman2021oa} mapped the global landscape of Diamond (no-fee, no-APC) open-access journals, providing the external baseline against which the Diamond OA advantage identified in Section \ref{sec:diamond} can be situated.

What distinguishes the present study from this body of work is scale and integration rather than any single method: prior studies typically examine comparatively small samples in terms of journals \cite{Acciai2023Estimating} or countries \cite{Oh2026Data}, 
a short time window \cite{Lawrimore2026Who}, 
or only a subset of FAIR dimensions \cite{Sinha2023Findability}, or they 
use AI/LLM-based classification on a scale this study's transparent rule-based approach does not attempt \cite{wong2026evaluating}.
Here, the same 6,874,176-article PMC-OA corpus (2011--2025) is scored by a single, transparent DAS-classification rule set, cross-validated against F-UJI's independent automated assessment, audited for JATS4R machine-readability, and stratified simultaneously by discipline, funder, geography, open-access tier, and inferred author gender, allowing the ten-year trajectory of FAIR uptake to be decomposed by \emph{who} adopted it, \emph{how} they documented it, and whether that documentation is actually machine-actionable, within one consistent corpus and scoring scheme.

\section{Methods}
\label{sec:methods}
\subsection{Study Design and Conceptual Framework}
\label{sec:studydesign}
This study adopts a longitudinal, corpus-based observational design to characterise how data-sharing practice in the biomedical and life-science literature evolved across a fifteen-year window (2011--2025) relative to the publication of the FAIR Principles \cite{wilkinson2016fair}. The year 2016 serves as the principal inflection point: articles published in the five years immediately preceding it (2011--2015) define the pre-FAIR baseline, and articles from 2016 onward are assessed for evidence of behavioural change aligned with Findability, Accessibility, Interoperability, and Reusability. The study also interrogates whether observed shifts are better attributed to concurrent policy drivers, namely journal data-availability mandates, funder open-data requirements, and the JATS4R markup recommendations, than to the FAIR Principles themselves. The correlational and before/after evidence presented here does not constitute claims about causality.

\subsection{Data Source and Corpus Construction}
\label{sec:datasource}
The primary data source is the PubMed Central Open Access Subset (PMC-OA), which distributes full-text articles as JATS (Journal Article Tag Suite) XML in a way that facilitates mining \cite{mietchen2013inconsistent}. Raw article XML was retrieved for publication years 2011--2025 via the PMC bulk-download / NCBI E-utilities interface, restricted to the open-access licence filter. Articles were included if they were parseable JATS records with a resolvable publication year; conference abstracts, errata, and other non-article record types are excluded at the parsing stage (Section \ref{sec:dataingestion}). The resulting master frame comprises 6,874,176 articles.

\subsection{Data Ingestion Pipeline and Database Schema}
\label{sec:dataingestion}
\begin{figure}[htbp]
\centering
\includegraphics[width=1.0\textwidth]{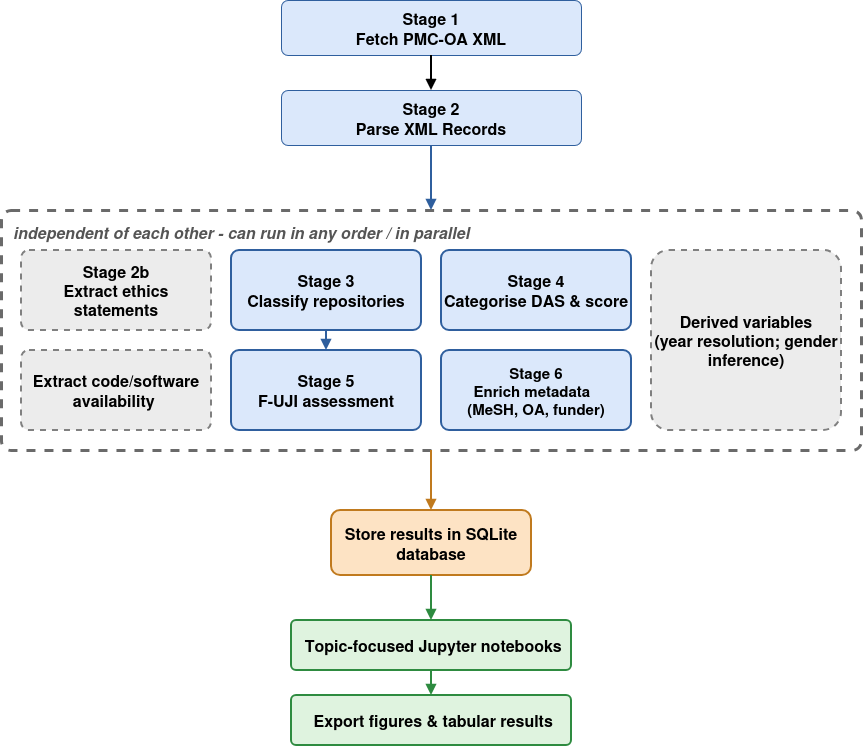}
\caption[]{Pipeline Architecture: From Raw PMC-OA XML to Analysis Notebooks. Stage 1 (Fetch) and Stage 2 (Parse) are strictly sequential. Everything inside the dashed region depends only on Stage 2's output, not on each other, and can run in any order or in parallel -- the one exception, Stage 5 (F-UJI assessment), additionally requires Stage 3's repository-tier classification and is drawn as its local dependent. 
The two small utility scripts that derive \texttt{article.year} (Section~\ref{sec:dataingestion}) and the offline gender-inference table (Section~\ref{sec:notebook}) are grouped together as ``Derived variables'' since neither persists a pipeline-numbered stage of its own.
We are exploring a range of variations to this basic workflow for future work and invite collaboration around that.
}
\label{fig:pipeline_overview}
\end{figure}

Article XML is processed by a sequence of numbered, purpose-built pipeline stages that populate a single normalised SQLite database. Each stage reads only the tables the previous stage(s) wrote and adds its own; every stage is idempotent and resumable, so an interrupted run can continue without duplication, and already-processed articles or URLs are simply skipped on re-run. The stages are:
\begin{enumerate}
  \item \textbf{Fetch} (Stage 1): retrieves one XML file per publication year from the PMC-OA bulk service via NCBI E-utilities, with rate-limiting and resumable per-year downloads.
  \item \textbf{Parse} (Stage 2): parses JATS XML into 12 relational tables: \texttt{journal}, \texttt{article}, \texttt{author}, \texttt{affiliation}, \texttt{article\_author}, \texttt{author\_affiliation}, \texttt{keyword}, \texttt{article\_keyword}, \texttt{data\_availability}, \texttt{da\_link}, \texttt{supplementary\_material}, and \texttt{accession\_number}. Non-article record types (conference abstracts, errata, corrections) are filtered out at this stage.
  \item \textbf{Ethics extraction} (Stage 2b): structurally parallel to Stage 2, populates \texttt{ethics\_statement} and \texttt{ethics\_sub\_type} (Section \ref{sec:extractionethics}).
  \item \textbf{Classify repositories} (Stage 3): classifies every URL in \texttt{da\_link} and writes \texttt{repo\_classification} (Section \ref{sec:repoclassification}).
  \item \textbf{Categorise DAS} (Stage 4): assigns each article's data-sharing category and data-sharing score, written to \texttt{das\_category} (Section \ref{sec:dascategory}).
  \item \textbf{F-UJI assessment} (Stage 5, with a dedicated retry pass for failed requests): submits URLs from repositories this pipeline has pre-classified as gold/silver tier\footnote{The terms``gold'' and ``silver'' stand for this pipeline's own repository-infrastructure tiers, defined in Section~\ref{sec:repoclassification} (persistent identifiers, open metadata, long-term preservation). They are used here only to select which data source URLs to submit to F-UJI. F-UJI itself does not use this terminology, and the tiers are unrelated to the Gold open access/subscription tiers of Section~\ref{sec:discipline} based on open-access terminology \cite{alperin_2026_21934115}. To minimize the potential for confusion, we will use lowercase ``gold''/ ``silver'' in the context of F-UJI assessment and uppercase ``Gold''/ ``Diamond'' in the context of open access throughout this manuscript.
  } to a self-hosted F-UJI instance,
  writing \texttt{fuji\_assessment}, \texttt{fuji\_metric}, and \texttt{fuji\_article\_link} (Section \ref{sec:automatedassessment}).
  \item \textbf{Enrich metadata} (Stage 6): adds discipline (\texttt{article\_mesh}, \texttt{article\_discipline}), open-access status (\texttt{journal\_oa}), and funder attribution (\texttt{article\_funder}) (Section \ref{sec:discipline}).
\end{enumerate}
Figure~\ref{fig:pipeline_overview} summarises this architecture end to end, from raw XML retrieval through to the figures and tabular exports reported in Section~\ref{sec:results}. 

\subsection{Extraction of Data Availability Statements}
\label{sec:extractiondas}
DAS text is extracted by a six-location priority cascade applied to each article's \texttt{<back>} and \texttt{<body>} elements, in the following order: (1) a legacy \texttt{<data-availability>} element; (2) a \texttt{<sec>} whose \texttt{sec-type} attribute matches a DAS-indicating pattern (e.g. \texttt{data-availability}, \texttt{availability}, \texttt{data-access}); (3) a \texttt{<sec>} with no \texttt{sec-type} whose \texttt{<title>} text matches the same pattern (e.g. "Data Availability", "Availability of data"); (4) a \texttt{<custom-meta>} element whose \texttt{<meta-name>} matches; (5) a \texttt{<notes>} element whose \texttt{notes-type} matches; (6) a footnote (\texttt{<fn-group>/<fn>}) whose text matches. The first match wins, and the full text and structural provenance (\texttt{source}, e.g. \texttt{sec[@sec-type='data-availability']} or \texttt{sec/title='Data Availability'}) are stored verbatim in \texttt{data\_availability}; this provenance string is the sole input to the JATS4R compliance classifier applied later (Section \ref{sec:JATS4R}). URLs embedded in the DAS text are extracted in parallel into \texttt{da\_link}.

\subsection{Repository Classification}
\label{sec:repoclassification}
The repository-classification stage (Stage 3) matches every URL in \texttt{da\_link} against a curated knowledge base of 81 domain/path regex patterns spanning nine repository types: general-purpose (e.g. Zenodo, Figshare, Dryad, Dataverse), domain-specific (e.g. NCBI GEO, SRA, GenBank, UniProt, PDB), code repositories (GitHub, GitLab), journal/publisher platforms, institutional repositories, cloud storage (Dropbox, Google Drive), preprint servers, "request-only" flagged links, and an unmatched "other" catch-all. Each matched entry carries a curated repository infrastructure tier: gold (typically offers a persistent identifier, open metadata, long-term preservation, e.g. Zenodo, GEO, SRA), silver (typically offers most of the above with some limitation, e.g. GitHub, OSF), or bronze (limited infrastructure support, e.g. journal supplementary portals, cloud storage), plus open-access and persistent-identifier flags. This tier is a static, repository-type-level heuristic, not a per-dataset FAIR assessment of what was actually deposited; Section~\ref{sec:automatedassessment} covers the latter via F-UJI on a gold/silver-tier subset. Unmatched URLs are tier \texttt{unknown}.

\subsection{DAS Categorisation and Proxy FAIR Scoring}
\label{sec:dascategory}
The DAS-categorisation stage (Stage 4) assigns each article to exactly one of nine mutually exclusive categories via rule-based regex matching on the DAS text (or, for articles with no DAS, on the presence of supplementary materials or inline accession numbers): \textit{repository\_named} (DAS names a specific repository with a link or ID), 
\textit{accession\_provided} (bare database accession number given, no named repository), \textit{open\_unspecified} (DAS states data are open but names no location), \textit{supplementary\_only} (data exist only as article supplementary files), \textit{available\_on\_request} (DAS states data available only on request), \textit{restricted} (DAS explicitly restricts access, e.g.\ for privacy/ethics reasons), \textit{no\_statement} (no DAS text found anywhere in the article), \textit{no\_das\_but\_supp} (no DAS, but supplementary materials are present), and \textit{no\_das\_but\_accession} (no DAS, but an inline accession number is present). A data-sharing score (0--10) is then computed deterministically from the category plus two capped bonuses, with an explicit per-category ceiling:

\begin{center}
\begin{tabular}{@{}lccl@{}}
\toprule
Category & Base & Bonuses available & Max \\
\midrule
repository\_named & 8 & licence +1, DOI +1 & 10 \\
accession\_provided & 7 & licence +1, DOI +1 & 9 \\
no\_das\_but\_accession & 6 & licence +1 & 7 \\
open\_unspecified & 4 & URL +1, licence +1 & 6 \\
supplementary\_only & 3 & DOI +1 & 4 \\
restricted & 2 & --- & 2 \\
no\_das\_but\_supp & 2 & --- & 2 \\
available\_on\_request & 1 & --- & 1 \\
no\_statement & 0 & --- & 0 \\
\bottomrule
\end{tabular}
\end{center}
Alongside the category and score, five secondary boolean/count signals are stored per article (\texttt{mentions\_license}, \texttt{mentions\_doi}, \texttt{mentions\_github}, \texttt{has\_url}, \texttt{has\_accession}), together with DAS \texttt{word\_count} and \texttt{sentence\_count}. The score is intentionally simple and transparent; it is a corpus-wide, queryable proxy and is not a substitute for the independent F-UJI assessment described next.

\subsection{Automated FAIR Assessment with F-UJI}
\label{sec:automatedassessment}
A subset of dataset URLs from gold- and silver-tier repositories (Section \ref{sec:repoclassification}) was independently assessed with the F-UJI automated FAIR-evaluation tool \cite{devaraju2021an}, self-hosted as a Docker container (\texttt{ghcr.io/pangaea-data-publisher/fuji}) and queried via its REST API in the F-UJI assessment stage (Stage 5, with a dedicated retry pass for failed requests). F-UJI evaluates 16 metrics across the four FAIR dimensions against a dataset's landing page; results are stored per assessed URL in \texttt{fuji\_assessment} and per metric in \texttt{fuji\_metric}. URLs were deduplicated before submission, since many articles cite the same dataset, and each unique assessment is linked back to every citing article via \texttt{fuji\_article\_link}, so a single F-UJI call can inform many articles' records without redundant API calls.

\subsection{Extraction of Ethics Statements}
\label{sec:extractionethics}
Ethics statements are extracted by the ethics-extraction stage (Stage 2b), structurally parallel to the DAS extractor, using a six-level priority cascade over each article's \texttt{<back>}, \texttt{<body>}, and \texttt{<front>} elements: (1) \texttt{<sec>} with an ethics-matching \texttt{sec-type}; (2) \texttt{<sec>} with no \texttt{sec-type} but an ethics-matching \texttt{<title>}; (3) \texttt{<notes>} with an ethics-matching \texttt{notes-type}; (4) \texttt{<fn>} with an ethics-matching \texttt{fn-type}; (5) \texttt{<p>} with an ethics-matching \texttt{content-type}; (6) \texttt{<named-content>} with an ethics-matching \texttt{content-type}. The first match wins, and the raw text plus structural provenance are stored in \texttt{ethics\_statement}. Nested sub-sections and typed inline content within the matched statement are additionally recorded as granular sub-type signals in \texttt{ethics\_sub\_type} (e.g. IRB approval, informed consent, animal/human-subjects flags), extracted the same way regardless of which cascade level matched the parent statement.

A further utility applies the same six-level cascade to standalone Code/Software Availability sections, written to \texttt{code\_availability} and \texttt{code\_link}. This closes a gap in the DAS extractor above (Section~\ref{sec:extractiondas}): its \texttt{sec-type}/title matching requires the word ``data'' to appear, so a section titled only ``Code Availability'' or ``Software Availability'' -- common from publishers such as PLOS, the Nature family, and eLife, especially from $\sim$2019 onward -- is otherwise missed entirely. The overlap with \texttt{data\_availability} for combined ``Data and Code Availability'' sections is deliberate rather than a duplication bug: such a section is captured by both extractors, and an explicit flag records whether a code statement was standalone or only embedded within a combined DAS, so downstream analysis can distinguish the two.

\subsection{Discipline, Open Access, and Funder Enrichment}
\label{sec:discipline}
The metadata-enrichment stage (Stage 6) adds three stratification variables used throughout the Results:
\begin{itemize}
  \item \textbf{Discipline.} Medical Subject Headings (MeSH) terms are fetched per article from PubMed E-utilities and mapped to one of 12 discipline buckets (genomics, structural biology, clinical, neuroscience, ecology, chemistry, bioinformatics, public health, pharmacology, immunology, microbiology, plant biology) via a curated regex map on the MeSH term text, with major-topic headings weighted 3$\times$ over non-major headings when an article carries more than one candidate discipline; articles with no informative MeSH mapping fall into a residual \textit{general} bucket, giving 13 discipline categories in total. Written to \texttt{article\_mesh} (per-term) and \texttt{article\_discipline} (one resolved discipline per article).
  \item \textbf{Open access status.} A per-journal classification into \texttt{gold\_oa} / \texttt{likely\_subscription} / \texttt{not\_listed\_in\_doaj} / \texttt{unknown} is assigned by first checking the publisher name against curated lists of known fully-OA publishers (e.g. PLOS, BioMed Central, Frontiers, MDPI, eLife $\to$ \texttt{gold\_oa}) and known subscription publishers (e.g. Elsevier, Springer, Wiley, Nature $\to$ \texttt{likely\_subscription}, a publisher-level heuristic only, since these publishers may also carry fully-OA journals); where neither heuristic matches, the journal's print and online ISSN are each queried against the DOAJ API \cite{doaj}, giving \texttt{gold\_oa} on a match and \texttt{not\_listed\_in\_doaj} (explicitly \emph{not} evidence of subscription status, since DOAJ is a directory of qualifying OA journals rather than a complete registry of subscription/hybrid titles) when both ISSNs were queried and neither matched; \texttt{unknown} covers journals with no ISSN to query at all. The result is also a \emph{current} DOAJ snapshot, not the journal's access model at a given article's actual publication year. Written to \texttt{journal\_oa}. This pipeline-level classification is intentionally coarse; the finer Diamond/Gold/hybrid/subscription split used in several Results figures is a notebook-level refinement described in Section \ref{sec:notebook}.
  \item \textbf{Funder attribution.} Funder names and grant IDs are extracted by regex from \texttt{<funding-group>} XML read directly from the raw article XML 
  and normalised against a curated list of major funding agencies (e.g. NIH, Wellcome, ERC, UKRI, DFG, NSF, Gates Foundation), each tagged with its funder country. Written to \texttt{article\_funder}.
\end{itemize}

\subsection{JATS4R Compliance Classification}
\label{sec:JATS4R}
Unlike the enrichment steps above, compliance classification against the JATS4R (JATS for Reuse) recommendations is \emph{not} a pipeline stage: it is implemented directly in the DAS and ethics-statement analysis notebooks as a pure function over the verbatim \texttt{source} provenance string the pipeline already stored (Section \ref{sec:extractiondas}, Section \ref{sec:diamond}); no new extraction or re-parsing of the original XML is performed. 
JATS4R's DAS recommendation \cite{jats4r2018das} specifies \texttt{sec-type="data-availability"} as the only compliant value; its ethics-statement recommendation \cite{niso2020ethics} specifies \texttt{sec-type="ethics-statement"}. 
The notebook classifier maps each \texttt{source} string to one of five tiers: \textbf{compliant} (exact match); \textbf{near-miss} (one of 16 named wrong-case/misspecified variants that JATS4R explicitly flags, e.g. \texttt{"Data Availability Statement"}, \texttt{"data\_availability"}); \textbf{title-only} (found only via heading text, no \texttt{sec-type} attribute at all); \textbf{non-standard} (a publisher-specific \texttt{custom-meta}, \texttt{notes}, or footnote element); and \textbf{no statement}. 
Because the pipeline records the publisher's own attribute value verbatim rather than a parsed/normalised category, this yields ground-truth, per-article compliance data rather than a sample-based estimate.

\subsection{Notebook-Level Derived Measures}
\label{sec:notebook}
Several measures reported in the Results are computed entirely within individual analysis notebooks, on top of the pipeline tables above, rather than being pipeline stages or persisted database columns in their own right:
\begin{itemize}
  \item \textbf{Diamond/Gold/hybrid/subscription OA tiers}: a dedicated analysis notebook refines the pipeline's coarse \texttt{journal\_oa} status by matching journal ISSNs against a curated Diamond OA list (drawn from the cOAlition~S 2021 Diamond OA study \cite{bosman2021oa}, a DOAJ APC=0 query, and manual verification against each publisher's own website) 
  and by publisher/title pattern matching, distinguishing no-fee Diamond OA from APC-funded Gold OA within what the pipeline had initially labeled \texttt{gold\_oa} (and, as corrected in Section \ref{sec:discipline} above, mapping \texttt{likely\_subscription} into this finer tier set's \texttt{subscription} label).
  \item \textbf{Author gender}: \emph{name-based gender inference}, not a measurement of author gender identity, computed offline in a standalone pipeline utility for every distinct given name in the corpus using a two-strategy classifier: the \texttt{names\_dataset} global name-gender probability database as primary source, with the rule-based \texttt{gender\_guesser} package as fallback, joined into the analysis notebooks by given name. 
  A name is classified \textit{female} or \textit{male} only if one gender's probability reaches an 0.80 threshold; bare initials and names below threshold are kept as \textit{ambiguous}/\textit{unknown} and excluded from binary comparisons. This method cannot identify non-binary authors, uses global rather than country-specific name distributions (so it is less reliable for East Asian, South Asian, and African naming traditions, where the same name can carry a different gender association than in the Western-skewed reference data. 
  Benchmarked error rates for the best-performing commercial tools range from 2.8\% on European names to 17.6\% on Asian names \cite{santamaria2018comparison}, and similarly degrade on biomedical-literature author names specifically \cite{sebo2021performance}), and have not been validated against any ground-truth sample of self-reported author gender for this corpus. Accordingly, every gender-stratified result in this manuscript should be read as an exploratory secondary signal rather than an established measurement (Section \ref{sec:persistentgender} in the Results discusses the effect on resolution completeness over time, and the confound this creates for the reported temporal trend). A country-aware alternative exists in principle: the World Gender-Name Dictionary pairs given names with country-specific gender distributions across 195 countries \cite{lax2021expanding}, but integrating it is left to future work.
  \item \textbf{Reproducibility signals R1--R6}: six independent regex detectors applied to each article's DAS text, namely code sharing (R1: GitHub/GitLab/code-hosting mentions), workflow-manager use (R2), environment specification (R3), version pinning (R4), interactive notebooks (R5), and pre-registration (R6), combined into a 0--6 composite score. These are notebook-defined patterns over the same \texttt{data\_availability.statement} text the pipeline extracted, not a separate extraction pass over the XML.
  \item \textbf{Mandate-compliance rubric}: a five-tier compliance label (fully compliant / compliant / partial / non-compliant / restrictive) is derived directly from each article's existing data-sharing score band (Section \ref{sec:dascategory}); no new signal is extracted. Funder-level and journal-level mandate milestone years (e.g. NIH 2003/2023, Wellcome 2011/2017, MRC 2013/2018, Horizon 2020 ORD Pilot 2017) are a curated reference table drawn from the funders' own published policies, not automatically detected from the corpus.
\end{itemize}

\subsection{Analytical Notebooks}
\label{sec:analytical}
All statistical analysis and figure generation were performed in 14 topic-focused Jupyter notebooks (database overview, corpus overview, DAS categories, JATS4R compliance, ethics statements, notebook/code sharing, gender, Diamond OA, mandate compliance, reproducibility, digital-use-conditions/OAMI, link rot, geography, and journal profiles), each querying the shared SQLite database directly via \texttt{pandas.read\_sql\_query}. Figures and tabular exports are written to dedicated output directories in the code repository; no analysis code operates on the raw XML directly, as every notebook reads only from the database tables populated by the pipeline stages above.

\subsection{Statistical Methods}
Group comparisons of data-sharing scores use the Mann-Whitney U test
, given the scores' bounded, non-normal, category-driven distribution.
Associations between categorical variables (e.g. DAS category $\times$ period, compliance tier $\times$ discipline) are tested with Pearson's chi-squared test of independence, reported with Cram\'er's V as effect size where stated. Linear associations between two continuous variables (e.g. reproducibility score versus
data-sharing score, notebook rate versus journal data-sharing score) use Pearson and, where noted, Spearman correlation. For standard treatments of the nonparametric, categorical-association, and correlation methods used in this paragraph, see \citep{pagano2022biostatistics}.
Simple linear trends over time (e.g. the female first-authorship share) are fitted with ordinary least-squares regression (\texttt{scipy.stats.linregress}), reported with $r^2$. The "pre-FAIR" (2011--2015) vs "post-FAIR" (2017--2025) and analogous before/after comparisons used throughout the Results are period-mean or period-proportion comparisons across this fixed split, not a fitted segmented-regression or interrupted-time-series model.
Difference-in-differences estimates (e.g. the gender data-sharing-score gap) are computed as the observed difference of period means between groups, with a 95\% confidence interval obtained by bootstrap resampling \citep{efron2016computerage} (10,000 resamples).
All notebooks use \texttt{scipy.stats}.
Co-evolution claims (e.g. DAS/notebook/ethics adoption tracking together over time) rest on Pearson correlation of annual prevalence rates only, not on a causal test.

\subsection{Reproducibility and Data Availability}
\label{sec:reproducibility}
The full pipeline (\texttt{pipeline/}), all analysis notebooks (\texttt{notebooks/}), and the resulting figures and tabular exports (\texttt{figures/}, \texttt{results/}) are maintained in the study's code repository alongside the SQLite database that backs every notebook query, so that any number reported in this manuscript can be traced back to the query that produced it. The code repository itself, and the processed database are each archived separately with their own DOI and licence, detailed in full in the Data availability and Code availability statements under Declarations; in brief, the pipeline and notebooks are released as \citep{samuel2026zenodocode}, and the processed SQLite database as \citep{samuel2026zenododb}.

\section{Results}
\label{sec:results}

We analysed 6,874,176 articles from the PubMed Central Open Access Subset published between 2011 and 2025 to characterise how data sharing practice has evolved relative to the 2016 publication of the FAIR Principles. Each article was scored on a 0--10 proxy FAIR scale derived from its data availability statement (DAS) category and secondary quality signals; a subset of dataset links was additionally assessed with the independent, automated F-UJI tool. The ten themes below summarise the corpus-wide trajectory, the persistent gaps between formal compliance and genuine machine-actionability, and where FAIR practice varies most by discipline, geography, funder, gender, and reproducibility.

\subsection{Corpus Size and DAS Adoption Over Time (2011--2025)}
\label{sec:corpusgrowth}
\begin{figure}
\centering
\includegraphics[width=1.0\textwidth]{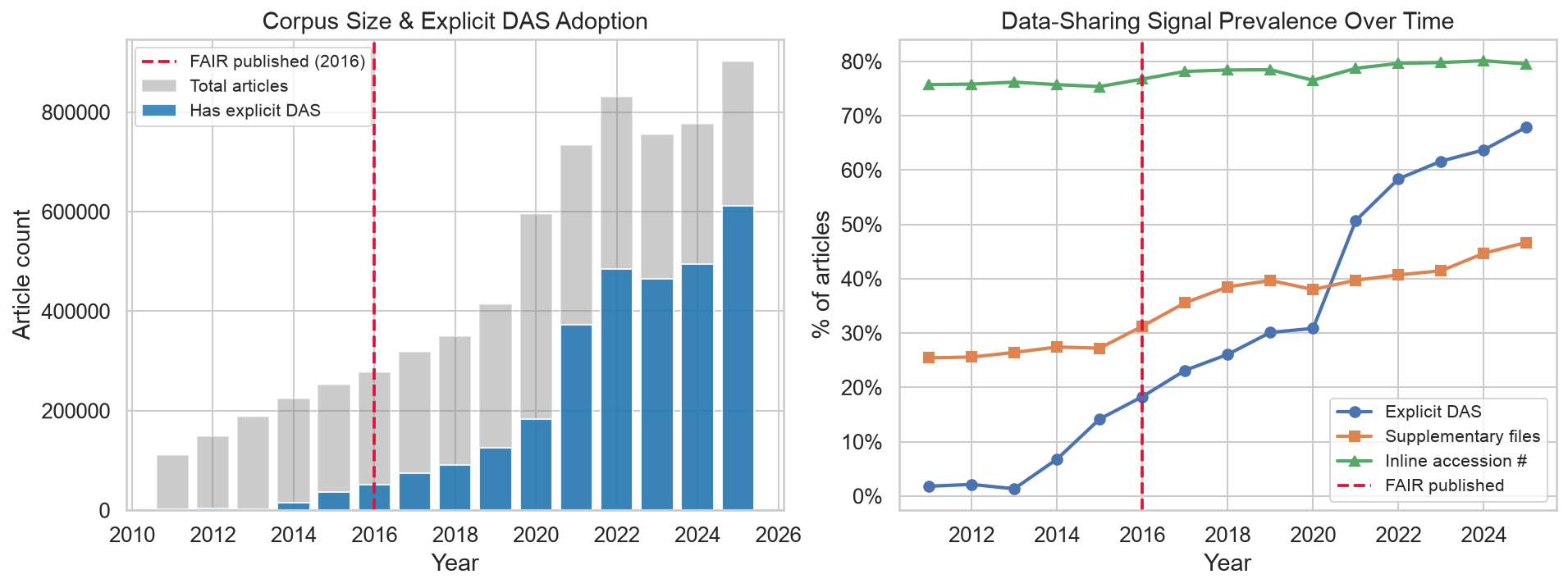}
\caption[]{Corpus Size \& Explicit DAS Adoption and Data-Sharing Signal Prevalence Over Time (2011--2025)}
\label{fig:fig_corpus_01_overview}
\end{figure}

The corpus grew from 110,347 articles per year in 2011 to 901,213 in 2025, peaking
above 830,000 in 2022~-- plausibly reflecting a broader pandemic-era publishing surge documented independently of this corpus \cite{rousseau2023publications,whitaker2025how}~-- before dipping to 754,942--776,984 in 2023--2024 and rebounding to its corpus-wide maximum of 901,213 in 2025 (Figure~\ref{fig:fig_corpus_01_overview}, left panel).
Against this growth, explicit DAS adoption underwent a dramatic transformation: from 1.8\% of articles in 2011, adoption accelerated via 6.8\% in 2014 (when the FAIR Principles went public on the FORCE11 website), 
crossed 18\% in 2016, reached 30\% by 2019--2020, and climbed to 67.9\% by 2025 (right panel). 
Supplementary-file prevalence rose more gently (25.4\%$\to$46.7\%), while inline accession numbers, the long-standing genomics proxy for data sharing, held steady around 75--80\% throughout (which is likely an overestimate, e.g. due to potential confusion between gene names and PDB identifiers). 
In absolute terms, the ``available on request'' anti-pattern grew from 77 articles (2011) to 173,474 (2025, a $>$2,250-fold increase off a near-zero base), while ``named repository'' grew from 132 to 94,697 ($\sim$720-fold) over the same span -- the anti-pattern's growth rate off its tiny 2011 base is over three times steeper than the best-practice pattern's, even though both are tracking the same broad post-2016 explosion in explicit DAS adoption (Figure~\ref{fig:fig_corpus_02_das_composition}, top panel).
The post-2016 surge in DAS adoption therefore represents a genuinely new documentation layer rather than a substitution for pre-existing sharing mechanisms. 

Because this study is based on the PMC Open Access Subset rather than all of PMC, two potential biases are worth noting explicitly. First, the OA Subset itself is not a fixed or representative slice of PMC: live NCBI \texttt{esearch} counts (queried directly against the same date-range and ``open access''[filter] term \texttt{pipeline/01\_fetch\_pmc.py} uses) show the OA Subset's share of all PMC articles published in a given year rising from 43.9\% (2011) to 91.2\% (2025), so the corpus studied here has grown from under half to nearly all of PMC's output, and any OA-Subset-specific bias (e.g. OA articles historically being more likely to carry an explicit DAS) should be shrinking over the study period rather than confounding the whole trend uniformly. This same 43.9\%$\to$91.2\% ratio also answers the related question of the annual ratio of OA-Subset articles to all PMC/PubMed articles, since it is computed from exactly that comparison. Second, the JATS XML tagging available through PMC is known to be internally inconsistent across publishers \cite{mietchen2013inconsistent}, which is precisely the gap
that JATS4R tries to address and that our JATS4R compliance analysis in Section~\ref{sec:machine} quantifies.

\subsection{Compositional Change in DAS Categories (2011--2025)
}
\label{sec:fromsilence}
\begin{figure}
\centering
\includegraphics[width=1.0\textwidth]{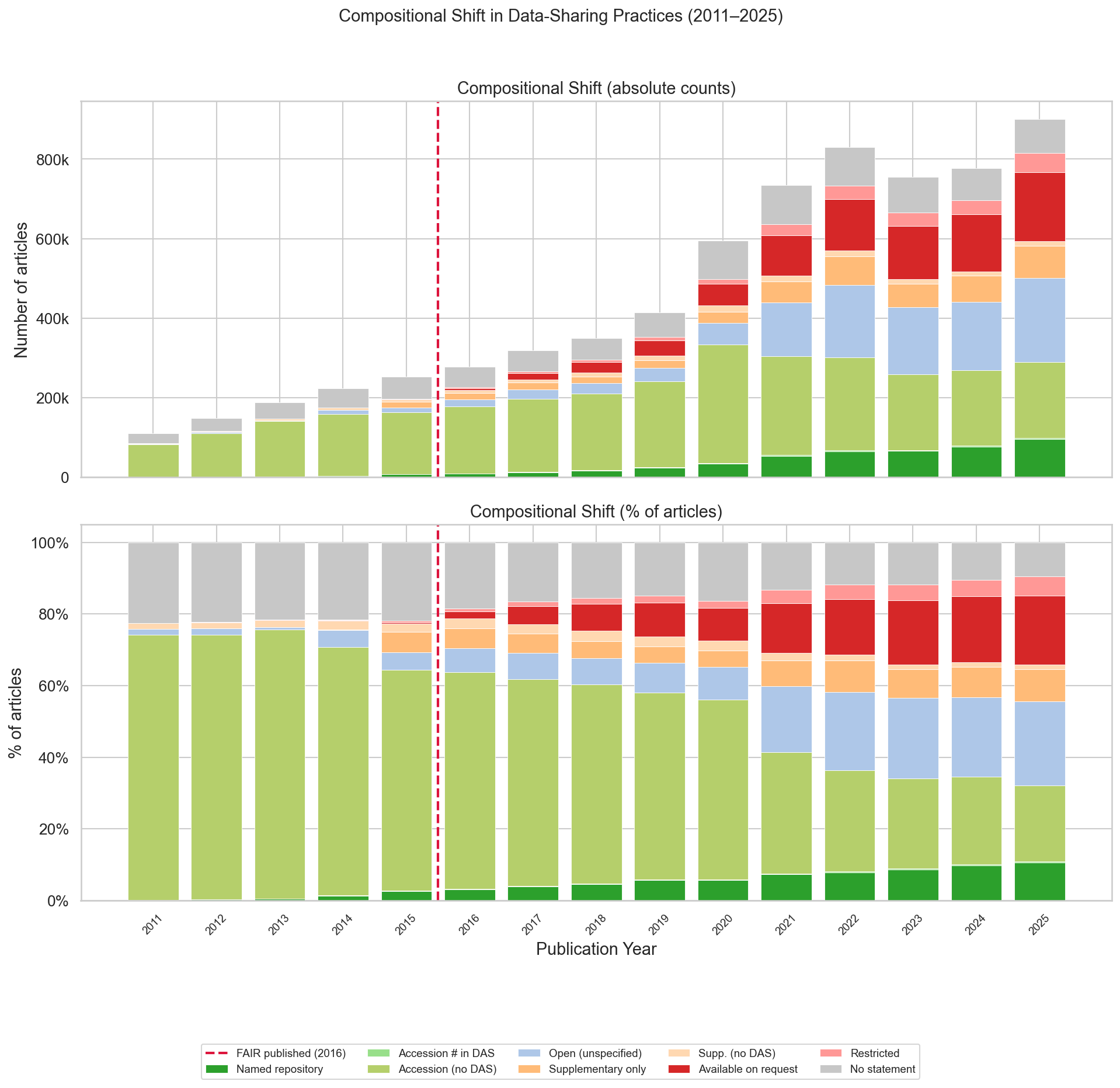}
\caption[]{
Compositional Shift in Data-Sharing Practices (2011--2025): Zooming in on the corpus description given in Fig.\ \ref{fig:fig_corpus_01_overview}, we present here
 absolute article counts (top) and \% composition (bottom) for the nine data-sharing categories and per-year totals, sharing one legend. %
The top panel shows every category grows in absolute terms, including both the best-practice ``named repository'' category (132$\to$94,697 articles, $\sim$720-fold) and the restrictive ``available on request'' anti-pattern (77$\to$173,474, $>$2,250-fold) -- the anti-pattern is not merely a shrinking share of a growing corpus, it is a larger absolute number of articles in 2025 than named-repository articles, even though the latter represent the recommended practice.
}
\label{fig:fig_corpus_02_das_composition}
\end{figure}

Beyond simple DAS adoption, the \emph{kind} of statement authors write changed markedly (Figure~\ref{fig:fig_corpus_02_das_composition}). In 2011, the corpus was dominated by bare accession numbers with no formal DAS (74.1\%) and a large ``no statement'' residual (22.5\%); named-repository citations (i.e.\ mentions of a known data repository inside a DAS) were essentially absent (0.1\%). 
By 2025, ``accession, no DAS'' had shrunk to 21.3\% and ``no statement'' to 9.6\%, while ``named repository''
grew to 10.5\%, ``open (unspecified)'' to 23.5\%, and the restrictive ``available on request'' anti-pattern rose from near zero to 19.3\% (a roughly 75-fold increase pooled across the pre-/post-2016 split; $\chi^2=1.47\times10^{5}$, $p<2.2\times10^{-16}$)\footnote{Every $p$-value this small or smaller is reported as $p<2.2\times10^{-16}$ (the standard bound once a $p$-value underflows double-precision floating point, rather than the literal and rather uninformative ``$p<0.001$''). \texttt{scipy.stats.chi2\_contingency} itself already returns exactly \texttt{0.0} for all of these, so $2.2\times10^{-16}$ is reported as an upper bound rather than the test's literal output} 
The best-practice ``named repository'' rate grew in parallel (1.2\%$\to$7.8\% pre/post-2016; $\chi^2=5.38\times10^{4}$, $p<2.2\times10^{-16}$): the two practices grew \emph{simultaneously} rather than one substituting for the other, and Shannon diversity of the nine-category mix rose from 0.71 to 1.90, confirming genuine diversification rather than a single dominant pattern shifting.

Naming a repository is also not sufficient for a genuinely high-infrastructure-tier destination: of DAS classified as \textit{repository\_named}, only 32.4\% point to a gold-tier repository (19.8\% silver, 47.6\% unknown); for \textit{accession\_provided} DAS the gold share is 20.6\%. Grouping ``available on request'', ``restricted'', and ``no statement'' into a combined restrictive bucket, this bucket rose from 22.3\% to 30.7\% of articles pre/post-2016 ($\chi^2=2.72\times10^{4}$, $p<2.2\times10^{-16}$): restrictive sharing grew in absolute terms even as best-practice sharing also grew. At journal level, the median post-2016 change in named-repository rate among 2,082 journals with sufficient pre/post data was only +1.1 percentage points, indicating corpus-wide gains are driven more by which articles and journals entered the corpus after 2016 than by existing journals changing their own norms.

\subsection{Data-Sharing Score Trend: Rising Adoption, Falling Quality}
\label{sec:thefairscoreparadox}
\begin{figure}
\centering
\includegraphics[width=1.0\textwidth]{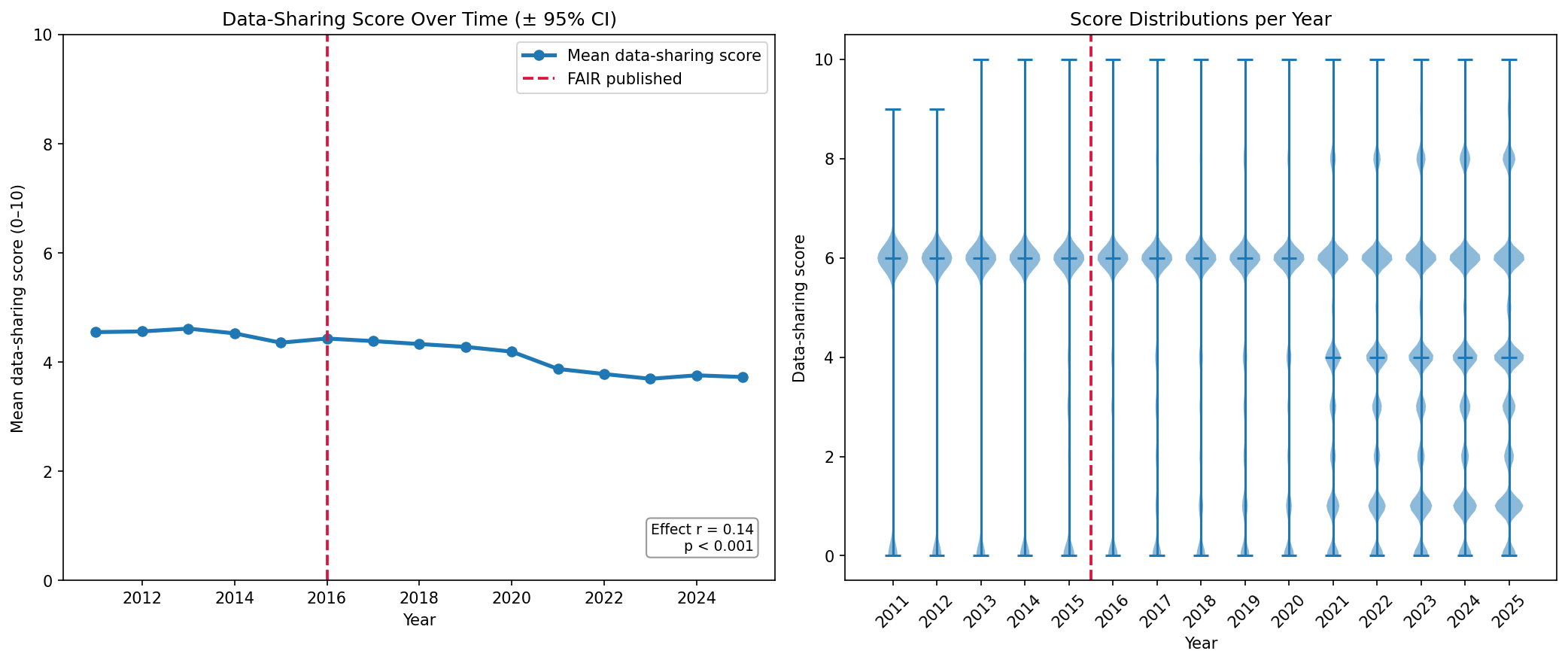}
\caption[]{Mean Data-Sharing Score Over Time (± 95\% CI) and Per-Year Score Distributions, 2011--2025
}
\label{fig:fig_corpus_04_fair_score_trend}
\end{figure}

\begin{figure}
\centering
\includegraphics[width=1.0\textwidth]{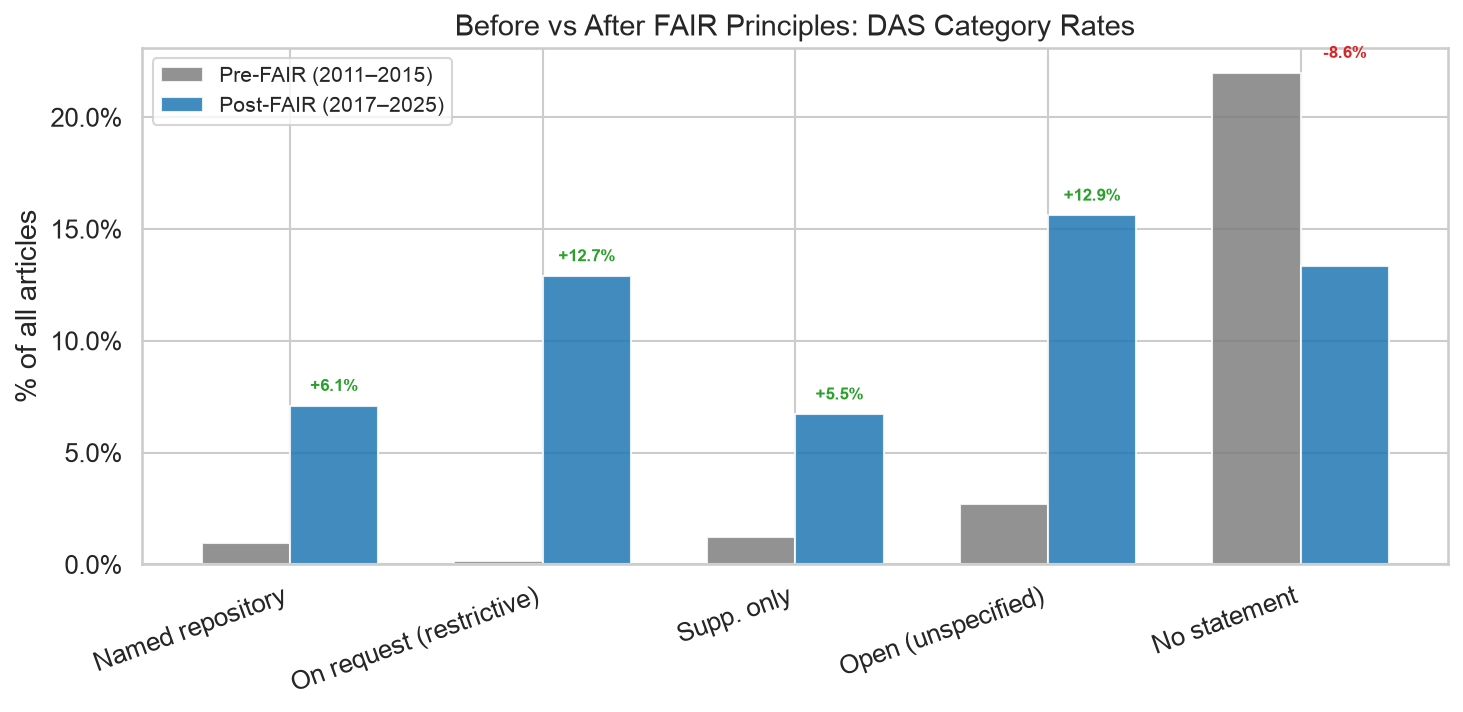}
\caption[]{Before vs After FAIR Principles: Summary Statistics Dashboard}
\label{fig:fig_corpus_11_before_after}
\end{figure}

Despite the rise in DAS adoption and named-repository usage -- echoing the pre-FAIR finding that data-sharing \emph{statements} do not reliably translate into data-sharing \emph{practice} \cite{vines2014the,federer2018data} -- the corpus's mean data-sharing score \emph{declined} over the study period, from 4.508 (2011--2015) to 3.956 (2017--2025; Mann-Whitney rank-biserial effect r=0.145). 
The year-by-year trend (Figure~\ref{fig:fig_corpus_04_fair_score_trend}, left panel) shows this is not a step change at 2016 but a gradual slide across the whole period, from a mean of 4.55 in 2011 to 3.73 by 2025, with the steepest single-year drop between 2020 and 2021 ($-$0.317, more than three times any other year-over-year change); per-year score distributions (right panel) show this is a genuine shift in the typical article, not only a thickening low-scoring tail: the per-year median itself steps down from 6 (2011--2020) to 4 (2021--2025). A structured pre-/post-FAIR comparison across the four headline corpus-level metrics confirms every trend is statistically significant and not an artefact of a shrinking sample (Figure~\ref{fig:fig_corpus_11_before_after}): explicit-DAS share rose over 8-fold, from 6.3\% to 51.1\% ($\chi^2=6.43\times10^{5}$, $p<2.2\times10^{-16}$, improved); ``available on request'' rose roughly 75-fold, from 0.2\% to 14.4\% ($\chi^2=1.47\times10^{5}$, $p<2.2\times10^{-16}$, worsened); named-repository share rose roughly 6.5-fold, from 1.2\% to 7.8\% ($\chi^2=5.38\times10^{4}$, $p<2.2\times10^{-16}$, improved); and mean data-sharing score fell 4.51$\to$3.92 (Mann-Whitney U=$3.01\times10^{12}$, $p<2.2\times10^{-16}$, declined).
The decline is best explained by compositional dilution: the growing share of low-scoring ``available on request'' and ``open (unspecified)'' articles documented above outweighs the growth of high-scoring named-repository citations, producing a category crossover in which ``no statement'' shrinks but is absorbed mainly by weaker explicit categories rather than by the strongest one.

\subsection{Repository Choice and F-UJI-Assessed FAIR Scores}
\label{sec:repositorychoice}
Ranked by raw citation count, the most-referenced destinations in DAS content are doi.org (116,421 links) and GitHub (110,967, silver tier), followed by a cluster of gold-tier, domain-specific repositories characteristic of molecular biology (NCBI GEO, GenBank, NCBI, OSF, BioProject, EMBL-EBI, NCBI SRA, and PDB),
with general-purpose repositories such as Zenodo and Figshare appearing further down the ranking. Code-repository and domain-specific-repository links both grow steadily across the period as explicit DAS sections spread beyond genomics into fields with less-established repository norms. Database accession numbers embedded directly in article text also grew substantially, with PDB, GEO, SRA, and GenBank remaining the dominant identifier families. Among secondary DAS quality signals, URL, DOI, and GitHub mentions all rose over time, while explicit licence mentions within DAS text remained rare throughout: reusability-relevant metadata remains uncommon even in articles with an explicit DAS.

\begin{figure}
\centering
\includegraphics[width=1.0\textwidth]{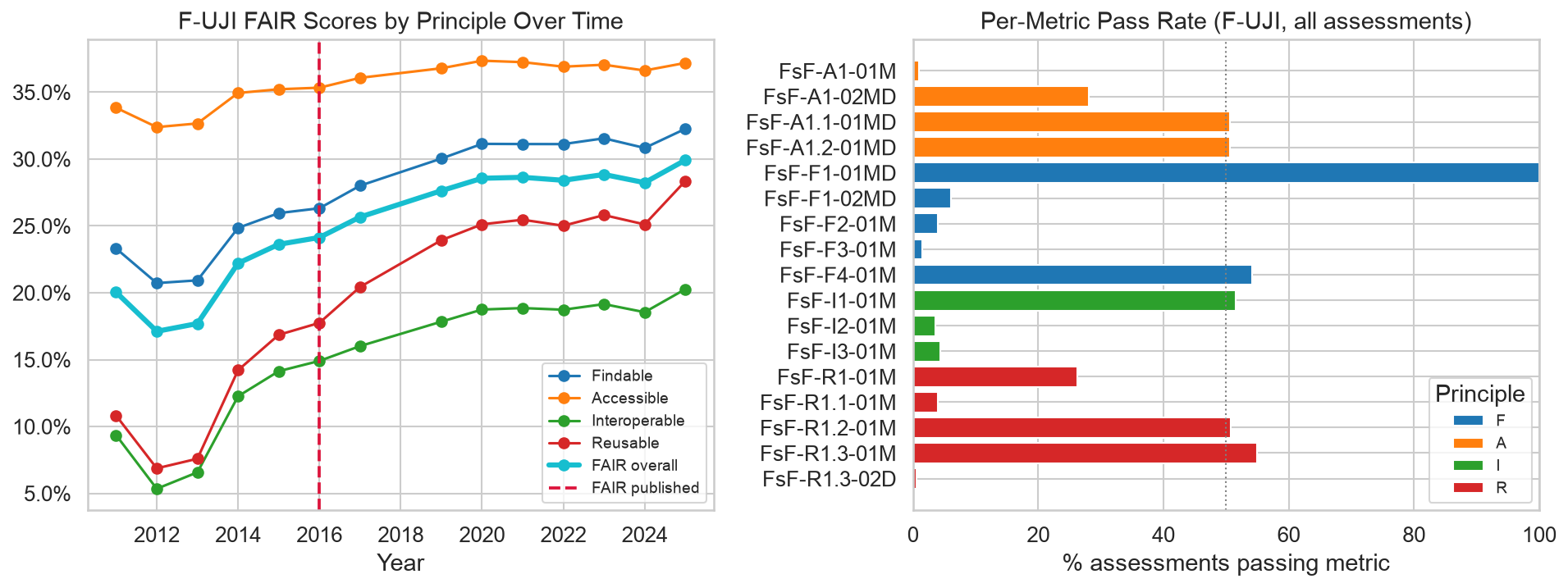}
\caption[]{F-UJI FAIR Scores by Access Protocol and Per-Metric Pass Rate (F-UJI, all assessments)}
\label{fig:fig_corpus_10_fuji_results}
\end{figure}

Independent, automated verification with the F-UJI tool on gold- and silver-tier dataset URLs corroborates this picture of uneven compliance across the four FAIR principles (Figure~\ref{fig:fig_corpus_10_fuji_results}). Breaking assessments down by access protocol, \texttt{https://} links, the large majority (192,672 assessments), achieve the highest mean F-UJI score (30.0; component scores F=32.5, A=37.2, I=20.4, R=28.3), while plain \texttt{http://} links score markedly lower (21.9; 16,011 assessments) and legacy \texttt{ftp://} links lowest of all (10.9; 1,191 assessments). Persistent, secure identification is thus associated with substantially higher automated FAIR compliance, yet a meaningful share of cited data links still uses legacy, lower-scoring access protocols.

\subsection{JATS4R Tagging Compliance}
\label{sec:machine}

\begin{figure}
\centering
\includegraphics[width=1.0\textwidth]{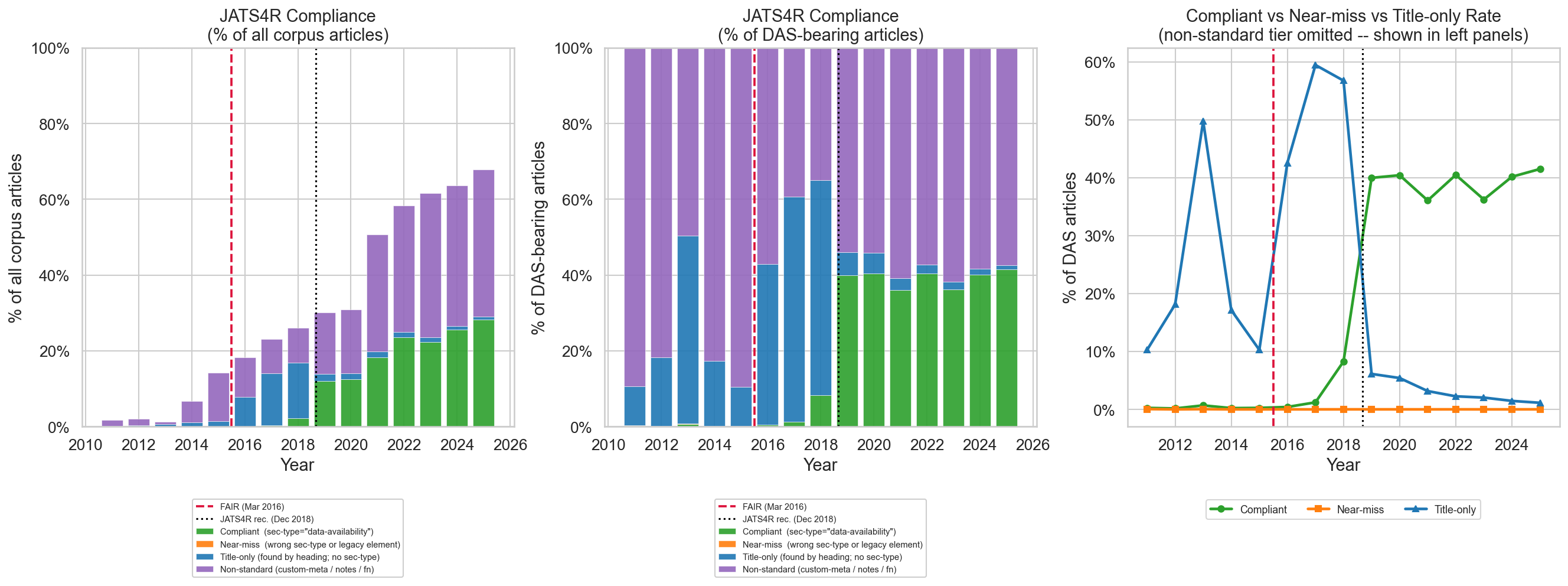}
\caption[]{JATS4R Compliance: Adoption Over Time Pre/Post  
the JATS4R Data Availability Statements Recommendation \cite{jats4r2018das}. 
Left and middle panels: compliance-tier composition as a \% of all corpus articles and as a \% of DAS-bearing articles respectively. Right panel: compliant, near-miss, and title-only rates plotted as separate lines (the fourth, largest tier, non-standard, is omitted here since it already appears in the other two panels, which is why these three lines do not sum to 100\%). 
}
\label{fig:fig_jats4r_04_over_time}
\end{figure}

Whether a DAS exists is only half the question; the JATS4R recommendation \cite{jats4r2018das} (published 28 December 2018) specifies exactly how DAS should be machine-tagged (\texttt{sec-type="data-availability"}, lowercase and hyphenated, no other value accepted). 
Corpus-wide, 43.8\% of articles carry a DAS of any kind, but only 36.0\% of those (15.8\% of the whole corpus) use the exact compliant tag; a further 2.8\% are ``title-only'' (heading text, no \texttt{sec-type}), 25.3\% use a publisher-specific non-standard element, and the remainder are near-miss variants among 45 distinct \texttt{sec-type} values actually observed in the corpus. 
Because the compliant value did not exist before the recommendation,
compliance is essentially zero before 2019 (0.2--0.7\% in 2011--2016)\footnote{JATS4R recommendations originate from a community process that makes drafts public before the official version is released. Those drafts, in turn, are often based on best practice already observed in the community, so it is technically possible for tagging to be compliant with a recommendation before the latter's formal release.} and then jumps to 40.0\% of DAS-bearing articles in 2019 (the first full post-recommendation year), settling in the 36--41\% band through 2025 (Figure~\ref{fig:fig_jats4r_04_over_time}): an uplift of +36.1 percentage points (compliant share of DAS-bearing articles, $\leq$2018 vs $>$2018) attributable to the recommendation itself rather than to the FAIR Principles three years earlier. 

JATS4R compliance is not a proxy for better data sharing practice: compliant-tier articles have a \emph{lower} mean data-sharing score (3.702) than title-only articles (4.006, $p<0.001$), indicating correct XML tagging and informative DAS content are governed by largely independent processes (one a publisher-side production choice, the other an author-side content decision), consistent with the broader reproducibility literature's caution \cite{munafo2017manifesto} against locating the whole burden of non-compliance in individual authors' behaviour.
JATS4R compliance also varies sharply by OA tier (Diamond OA highest at 62.8\%, Gold OA 
lowest at 21.8\%, subscription in between at 56.6\% )
and, of 1,157,013 structured data-citation links recorded, none carry the \texttt{@publication-type="data"} attribute JATS4R Rule 5 requires. The same gap appears, at an even lower baseline, for ethics-statement tagging: only 2.1\% of all articles use the exact compliant \texttt{sec-type="ethics-statement"} value (published 1 April 2020), rising to just 15.3\% of ethics-statement-bearing articles in the post-recommendation era (2020--2025). Articles with a compliant ethics tag are more likely (92.0\%) to also carry a DAS of some kind than articles with a non-compliant ethics tag (63.1\%), which fits with
XML tagging quality being a publisher-level property 
rather than an article-specific choice (Table~\ref{tab:jats4r_publisher}). 

\begin{table}[htbp]
\centering
\caption{JATS4R Compliance by Publisher: Top 10 publishers by article volume regardless of compliance rate, Top and Bottom 10 Publishers by compliance rate (among publishers with $\geq$50 articles; bottom list restricted to publishers with $\geq$20 DAS-bearing articles). $n$: number of articles by the publisher in our corpus; \%\,c: Percentage of DAS-bearing articles using the exact compliant \texttt{sec-type="data-availability"} tag, shown to two decimal places in the volume-sorted block to distinguish a genuinely tiny but nonzero rate (e.g.\ MDPI, 0.01\%, 74 of 602,812 DAS-bearing articles) from the literal 0.00\% of the bottom-compliance block; MDPI does not itself appear there because 41 other publishers (at $\geq$50 articles and $\geq$20 DAS-bearing articles) have exactly zero compliant articles and rank lower still.
}
\label{tab:jats4r_publisher}
\small
\begin{tabularx}{\textwidth}{@{}X r r@{}}
\toprule
Publisher & $n$ & \%\,c \\
\midrule
\multicolumn{3}{@{}l}{\textbf{Top 10 by article volume} (regardless of compliance rate)} \\
\cmidrule(r){1-3}
Multidisciplinary Digital Publishing Institute (MDPI) & 919,296 & 0.01 \\
BMC & 610,758 & 3.55 \\
Frontiers Media SA & 544,704 & 99.25 \\
Wiley & 394,804 & 97.42 \\
PLOS & 368,194 & 0.24 \\
Nature Publishing Group & 360,169 & 1.55 \\
Elsevier & 300,746 & 86.64 \\
Oxford University Press & 237,032 & 90.98 \\
Wolters Kluwer -- Medknow Publications & 136,113 & 59.96 \\
Cureus Inc. & 100,517 & 0.00 \\
\midrule
\multicolumn{3}{@{}l}{\textbf{Top 10 by compliance rate} (publishers with $\geq$50 articles)} \\
\cmidrule(r){1-3}
PAGEPress & 2,819 & 100.0 \\
Jaypee Brothers Medical Publishing (P) Ltd. & 2,674 & 100.0 \\
Yonsei University College of Medicine & 2,529 & 100.0 \\
SC Jurnalul pentru Medicina si Viata SRL & 2,138 & 100.0 \\
Association for Dental Sciences of the Republic of China & 1,984 & 100.0 \\
Scientific Foundation SPIROSKI & 1,805 & 100.0 \\
Royan Institute & 1,789 & 100.0 \\
Nepal Medical Association & 1,686 & 100.0 \\
California Institute of Technology & 1,597 & 100.0 \\
Korean Surgical Society & 1,553 & 100.0 \\
\midrule
\multicolumn{3}{@{}l}{\textbf{Bottom 10 by compliance rate} ($\geq$20 DAS-bearing articles)} \\
\cmidrule(r){1-3}
Cureus Inc. & 100,517 & 0.0 \\
Medicina Oral S.L & 3,707 & 0.0 \\
Korean Society of Anesthesiologists & 2,387 & 0.0 \\
Open Medicine & 1,782 & 0.0 \\
The Korean Neurosurgical Society & 1,759 & 0.0 \\
Shahid Sadoughi University of Medical Sciences and Health Services & 1,560 & 0.0 \\
Korea Disease Control and Prevention Agency & 1,379 & 0.0 \\
Indian Orthopaedic Association & 1,368 & 0.0 \\
Korean Society of Gynecologic Oncology and Colposcopy & 1,315 & 0.0 \\
Korean Society of Community Nutrition and the Korean Nutrition Society & 1,244 & 0.0 \\
\bottomrule
\end{tabularx}
\end{table}

Ethics statements are also linked to data sharing in a more complex way than volume alone suggests: articles carrying \emph{both} an ethics statement and a DAS have the \emph{lowest} mean data-sharing score of any presence/absence combination (3.375, versus 4.320 for neither), consistent with human-subjects research (which accounts for the majority of articles with ethics statements) more often falling into lower-scoring restrictive DAS categories, usually for privacy reasons.

\subsection{Data-Sharing Score by Discipline and Geography}
\label{sec:disciplinary}
\begin{figure}
\centering
\includegraphics[width=1.0\textwidth]{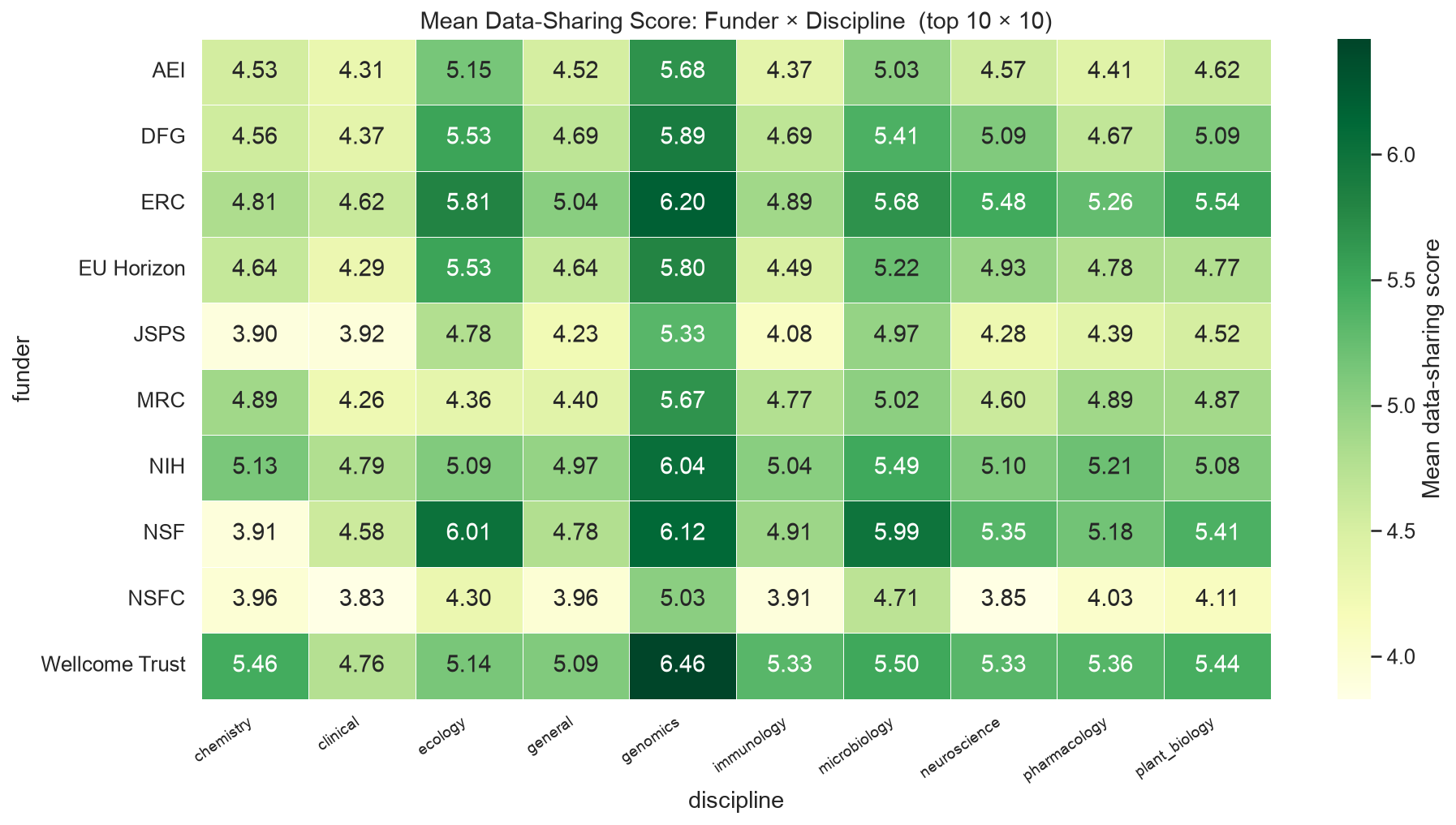}
\caption[]{Mean Data-Sharing Score: Funder $\times$ Discipline (top 10 funders $\times$ top 10 disciplines by volume). NASA and NIH score highest in almost every discipline cell they fund, while NSFC and JSPS %
trail even within the same disciplines. The funder effect documented in Section~\ref{sec:funder} holds up within discipline, not merely in aggregate.}
\label{fig:fig_geo_07_funder_discipline}
\end{figure}

\begin{figure}
\centering
\includegraphics[width=1.0\textwidth]{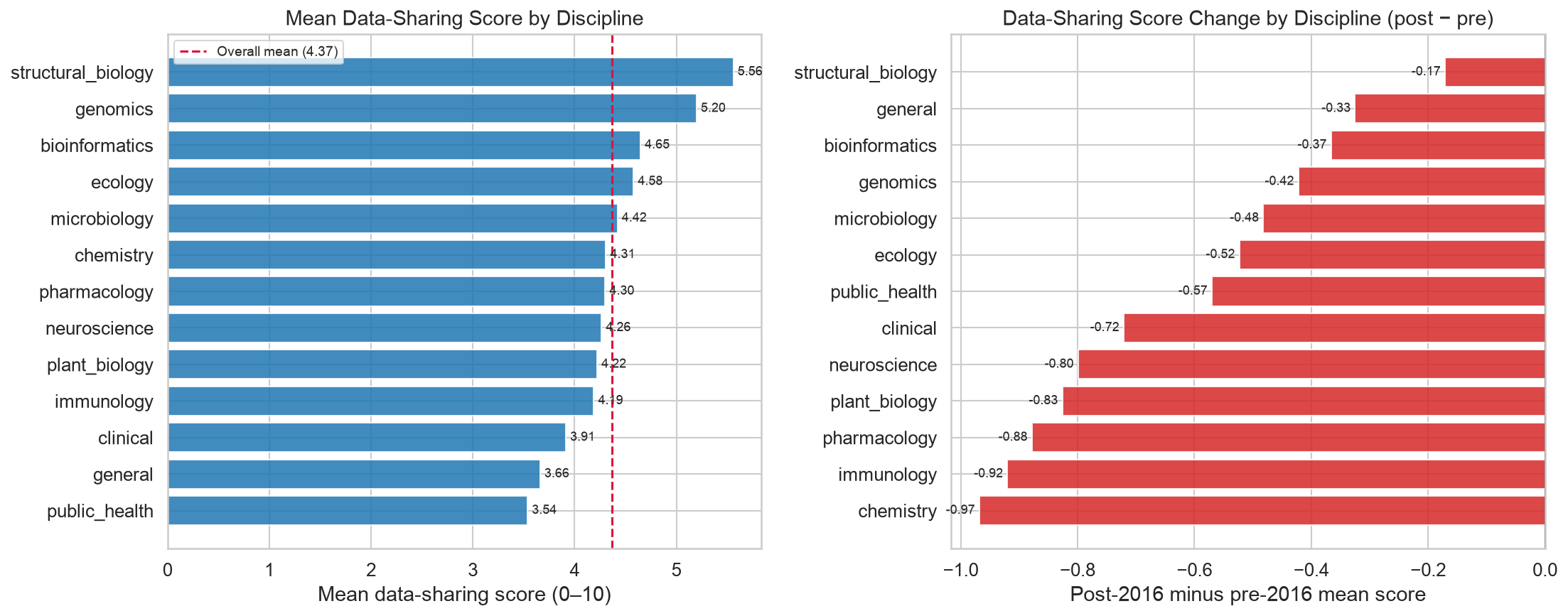}
\caption[]{Mean Data-Sharing Score and Data-Sharing Score Change, both by discipline (post - pre). %
}
\label{fig:fig_geo_01_discipline_scores}
\end{figure}

Mean data-sharing scores vary substantially by discipline, from 5.558 in structural biology and 5.196 in genomics down to 3.912 in clinical research and 3.536 in public health (Figure~\ref{fig:fig_geo_01_discipline_scores}, left panel). Comparing each discipline's post-2016 mean against its own pre-2016 mean, every one of the 13 disciplines declined (right panel):
structural biology shows by far the smallest decline ($-$0.171), while chemistry ($-$0.968), immunology ($-$0.920), and pharmacology ($-$0.878) declined the most despite high pre-2016 baselines. The same pattern holds when disciplines are derived from MeSH major headings instead of the coarser 13-category classification, and largely persists when cross-tabulated against funder:
NASA and NIH remain among the highest-scoring cells across almost every discipline they fund (Figure~\ref{fig:fig_geo_07_funder_discipline}).

\begin{figure}
\centering
\includegraphics[width=1.0\textwidth]{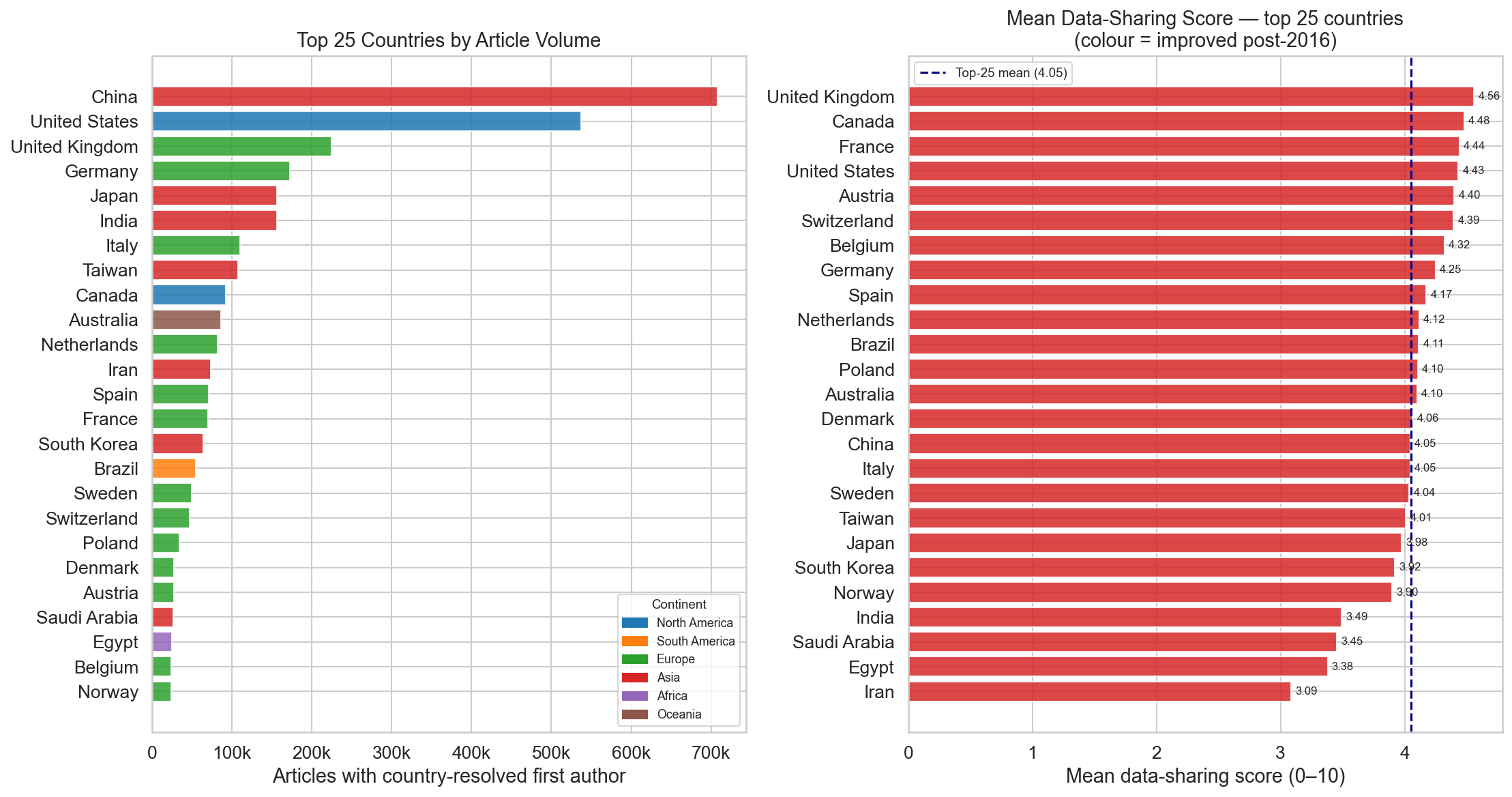}
\caption[]{Top 25 Countries by First-Author Article Volume (left; colour groups all 25 countries by continent -- North America: US, CA; South America: BR; Europe: GB, DE, IT, NL, ES, FR, SE, CH, PL, DK, AT, BE, NO; Asia: CN, JP, IN, TW, KR, IR, SA; Africa: EG; Oceania: AU -- 
purely to aid reading the bars, not an improved/declined signal) and Mean Data-Sharing Score for the same 25 countries (right; colour there does mean improved, green, vs declined, red, pre/post-2016).
China and the United States dominate by volume, but the United Kingdom, Canada, and France lead on mean score; every one of the top 25 declined from pre- to post-2016 regardless of volume or starting score.}
\label{fig:fig_geo_09_country_volume_score}
\end{figure}

By first-author affiliation 
(Figure~\ref{fig:fig_geo_09_country_volume_score}), China and the United States dwarf all other countries in volume, but the United Kingdom has the highest mean score among the top 25 producing countries (4.559), while China sits below the top-25 mean (4.051). 
The decline seen at discipline level is close to universal geographically: all top-25-by-volume countries declined from pre- to post-2016, with the steepest drops in Taiwan ($-$1.629) and South Korea ($-$1.335), and even substantial declines in the UK, Germany, China, and the US. 
The mean data-sharing score also decreases monotonically with World Bank income group, from 4.244 (high-income) to 2.962 (low-income), and low-income countries show the largest pre/post-2016 decline ($-$1.181) of any income group. Internationally co-authored articles score markedly higher throughout (4.394 vs 3.971 for domestic-only, $p<0.001$), one of the larger single-variable splits documented in this study, 
yet even the ten countries (out of all 219 resolved countries meeting this metric's own $\geq$30-articles-per-period bar, not only the top 25 by volume)
with the \emph{largest} improvement in high-scoring data sharing after 2016 all still show net declines, underscoring how broad-based the corpus-wide erosion is. The erosion is not quite exceptionless, however: among all 219 resolved countries, exactly one clears the same $\geq$50-articles-per-period bar used above with a genuine positive mean-score delta: Nepal (3.130 pre-2016 $\to$ 4.033 post-2016, $+0.902$, $n=4{,}990$).

\begin{figure}
\centering
\includegraphics[width=1.0\textwidth]{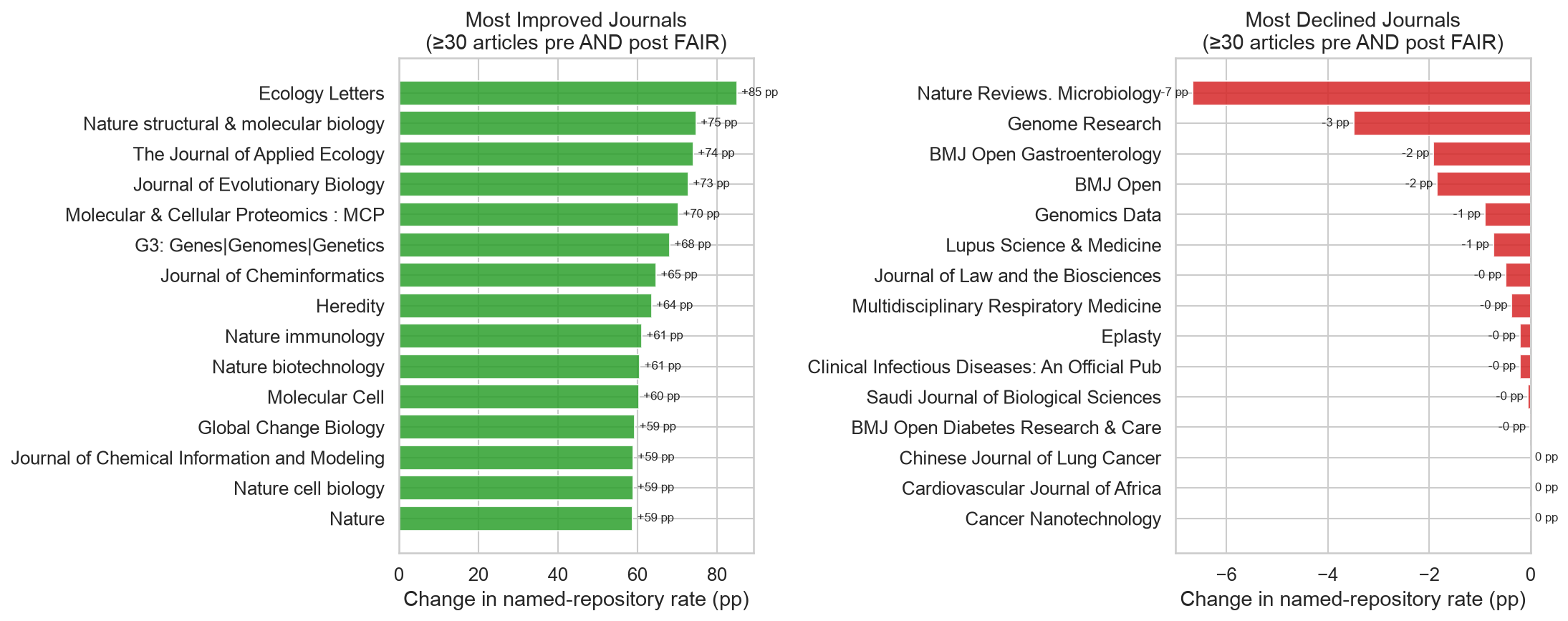}
\caption[]{Most Improved and Most Declined 
Journals by Change in Named-Repository Rate, Pre- vs Post-FAIR (journals with $\geq$30 articles on both sides of 2016; ``pp'' = percentage points, i.e.\ the raw change in named-repository rate, not a relative/percentage change of that rate). Ecology Letters, structural/molecular-biology journals (Nature structural \& molecular biology, Molecular \& Cellular Proteomics), and genomics journals (G3, Genome Medicine, Genome Biology) show the largest gains; the most-declined list is dominated by very small absolute shifts (at most $-$7pp) on an already-small base, consistent with Section \ref{sec:fromsilence}'s point that the corpus-wide decline is driven by corpus composition rather than by existing venues retreating from good practice.}
\label{fig:fig_dascat_10_trajectories}
\end{figure}

At the journal level, the decline is broad-based but not universal (cf.\ Fig. \ref{fig:fig_dascat_10_trajectories}): of the 2,082 journals with $\geq$30 articles on both sides of 2016 (Section~\ref{sec:fromsilence}), 754 (36.2\%) show a \emph{positive} pre/post data-sharing-score delta despite the corpus-wide mean falling. The largest gains cluster in ecology and chemistry informatics rather than in the highest-baseline disciplines: Journal of Applied Ecology ($+3.83$, $n_{\text{pre}}=49$, $n_{\text{post}}=43$), Journal of Cheminformatics ($+3.57$, $n=591\to939$), and Ecology Letters ($+3.34$, $n=78\to387$) are among the journals with the largest gains, alongside several journals whose post-2016 volume grew severalfold over the same window -- consistent with Section \ref{sec:fromsilence}'s point that corpus-wide gains are driven more by composition than by existing venues uniformly raising their own bar. 

\begin{figure}
\centering
\includegraphics[width=1.0\textwidth]{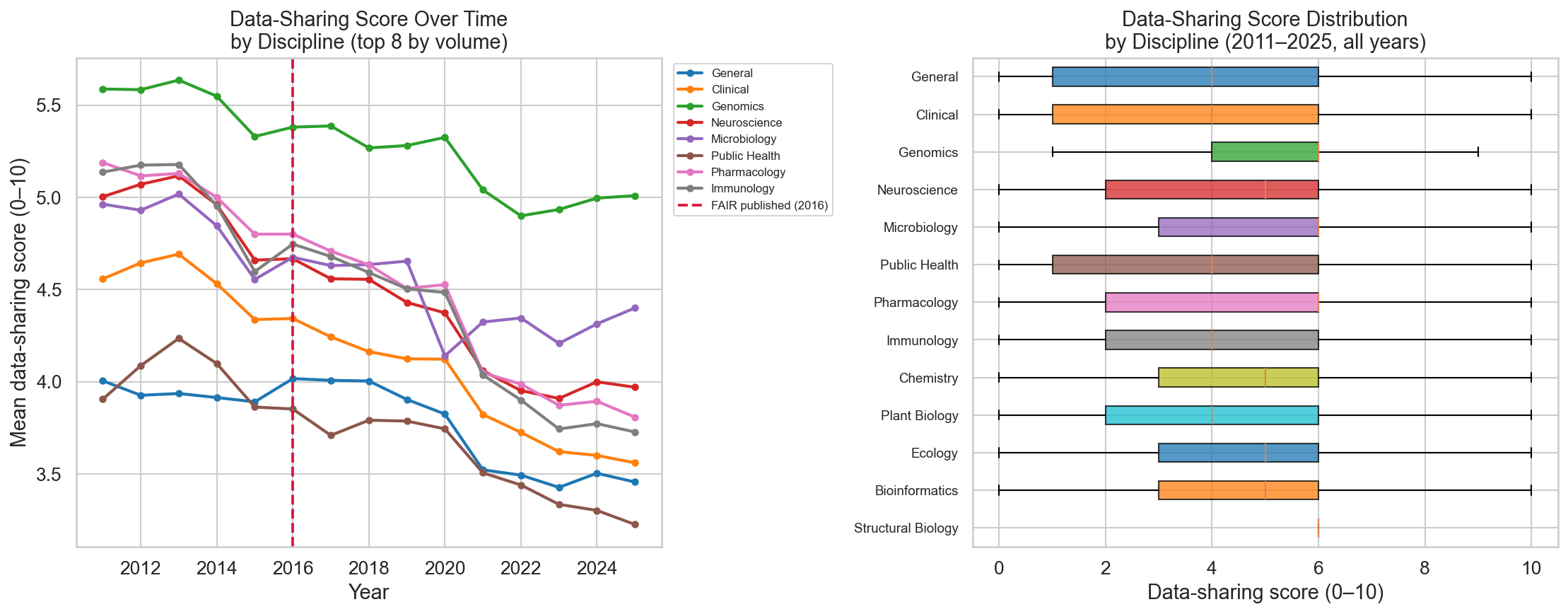}
\caption[]{Data-Sharing Score Over Time by Discipline (top 8 by volume; left) and Data-Sharing Score Distribution by Discipline, All Years Pooled (right; box = IQR, whiskers = 1.5$\times$IQR, median line). The left panel shows every discipline declining from a 2011--2013 peak, with genomics staying highest throughout and public health/general lowest; the right panel shows genomics's and structural biology's distributions are both narrower and shifted higher than clinical/public health's, reflecting the same gap already summarised by discipline means in Figure~\ref{fig:fig_geo_01_discipline_scores} but now showing the full spread, 
not just the mean.}
\label{fig:fig_geo_17_discipline_trend_dist}
\end{figure}

\subsection{Funder and Journal Mandate Compliance}
\label{sec:funder}
\begin{figure}
\centering
\includegraphics[width=1.0\textwidth]{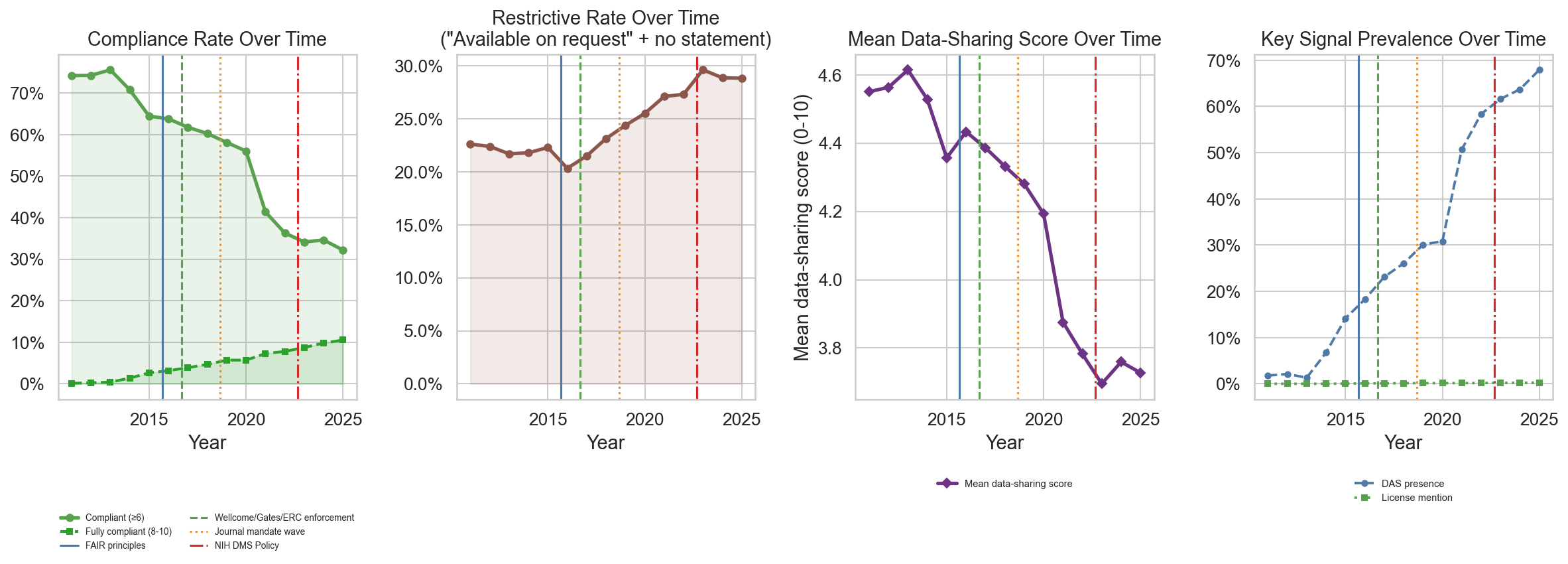}
\caption[]{Mandate Compliance Tier Trend Over Time, 2011--2025, in four single-metric panels: compliance rate (\% meeting the five-tier rubric's ``compliant'' threshold or above), restrictive rate (\% in the restrictive band), mean data-sharing score, and key-signal prevalence (licence, DOI, and named-repository mentions). The ``fully compliant'' share rises steadily throughout, while the broader ``compliant'' share and the mean score both decline from the early 2020s, the corpus-wide pattern described in the text above playing out tier by tier over time. 
}
\label{fig:fig_mandate_02_trend}
\end{figure}

Scoring every article against a five-tier mandate-compatibility rubric, 47.1\% of the corpus meets or exceeds the ``compliant'' threshold while 26.1\% falls in the restrictive band. Tracking compliance tiers over time reveals a striking, counter-intuitive trend (Figure~\ref{fig:fig_mandate_02_trend}): the ``fully compliant'' share rose steadily from 0.1\% (2011) to 10.5\% (2025), but the broad ``compliant'' share fell sharply from 74.1\% to 21.7\% as the ``partial'' tier grew from 1.5\% to 23.6\%: mandates appear to be pushing a small top tier toward full compliance while pushing a much larger share of the corpus down 
from ``compliant'' into ``partial'', net negative for the corpus median.
Put in absolute terms, this is not a shrinking tier: the number of compliant articles still grew from 81,748 (2011) to 195,210 (2025). But that 2.4-fold growth is considerably slower than the corpus's own 8.2-fold growth over the same period, and far slower still than the 719-fold growth of the fully-compliant tier and the 126-fold growth of the partial tier -- so the ``compliant'' tier is not disappearing, it is simply being vastly outpaced by growth concentrated at both the top and the newly-expanded middle of the rubric.

Among funders with known data-sharing mandates and sufficient volume (n$\geq$50 funder-article records), compliance rates span a comparatively narrow range: ERC (62.8\% compliant, mean score 5.30) and NASA (66.6\%, 5.91) lead, with NIH close behind (60.8\%, 5.23); AEI (49.4\%, 4.71) and MRC (49.9\%, 4.66) trail, though both remain above 30\% compliance.
Funder-level pre/post-strong-mandate-year deltas do not show a clear ``policy bend'' for any major funder: of the tracked funders, only BBSRC (2018 policy: $\Delta=+0.011$, $p<0.001$) and Wellcome Trust (2017 mandate: $\Delta=+0.007$, $p<0.001$) reach statistical significance in the positive direction, and both deltas are practically negligible; NIH's 2023 Data Management and Sharing Policy ($\Delta=-0.248$), NSF's 2023 policy ($\Delta=-0.374$), and MRC's 2018 policy ($\Delta=-0.575$) all show a \emph{decline} from pre- to post-mandate-year, none reaching significance in the positive direction.

This is consistent with the corpus-wide secular decline in mean data-sharing score documented in Section~\ref{sec:thefairscoreparadox} outpacing any mandate-specific improvement for articles published in these funders' recent, post-mandate years: a simple pre/post comparison cannot disentangle a within-funder policy effect from the corpus-wide trend. Repository choice matters even where data is shared: gold-tier repository links reach 97.0\% mandate compliance (mean score 7.9) versus only 26.0\% (4.7) for bronze-tier hosts. US-based funders (59.1\%) show only a modest aggregate-compliance edge over EU (57.5\%), UK (56.9\%), and German (55.2\%) funders, while Chinese funders lag more substantially (38.1\%). Finally, a four-criterion DAS completeness score (named repository + persistent identifier + licence + access condition, 0--4) rose monotonically across mandate eras, from 0.033 pre-mandate to 0.266 in the strong-mandate era (2023--2025): steady but slow progress toward a still-low ceiling (corpus-wide mean 0.185/4).

\subsection{Data-Sharing Score by Open-Access Tier}
\label{sec:diamond}
Gold OA journals showed a modest mean-score advantage over subscription journals (4.056 vs 3.913). 
Both tiers \emph{declined} from pre- to post-2016 (gold: $\Delta=-0.864$; subscription: $\Delta=-0.749$), and at similar rates.
A finer four-tier classification reveals a much larger effect the binary split obscures: Diamond OA (no author-facing charge, no reader-facing charge) has by far the highest mean score of any tier, 5.618, versus 4.056 for Gold OA and 3.913 for subscription (Mann-Whitney vs Gold: $p<0.001$, rank-biserial r=0.360; vs subscription: $p<0.001$, r=0.347).
Diamond OA also has the highest rate of high data-sharing scores (65.04\% vs 48.55\% subscription, 36.19\% Gold) and lowest restrictive-sharing rate (5.53\%), and leads every secondary quality signal examined.

\begin{figure}
\centering
\includegraphics[width=1.0\textwidth]{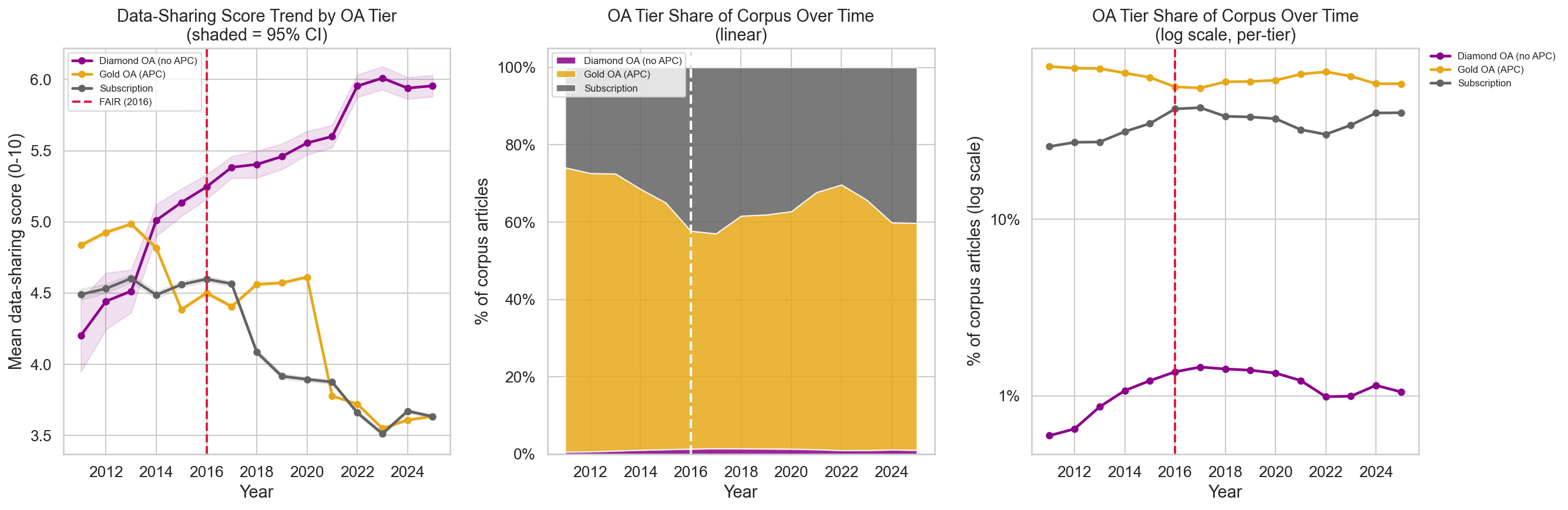}
\caption[]{Data-Sharing Score Trend by OA Tier, Pre vs Post 2016
Left: mean score by tier, shaded 95\% CI. Middle: tier share of the corpus, linear scale (stacked area).
Right: the same tier shares, log scale, plotted as separate lines rather than stacked.

}
\label{fig:fig_d06_trends}
\end{figure}

Tracked over time (Figure~\ref{fig:fig_d06_trends}), Diamond OA is the \emph{only} tier whose mean score \emph{improved} from pre- to post-2016 (4.833$\to$5.745, $\Delta=+0.911$), while Gold OA and subscription both declined substantially and at similar rates. 
Diamond OA's advantage is a post-2016 phenomenon, not a stable historical baseline, and it is now the only tier moving against the corpus-wide decline rather than one of two tiers eroding at different rates. As a focused case study, the two Diamond journals in chemistry/nanotechnology published by the Beilstein Institute (n=6,483 combined) score significantly above the corpus overall (4.322 vs 4.018) but significantly below the rest of the Diamond OA population (5.834); their above-average score is driven by chemistry's accession-number culture rather than richer DAS text, illustrating that Diamond OA's tier-level advantage is concentrated in a specific set of high-performing journals rather than uniform across all Diamond OA venues.

\subsection{Data-Sharing Score by Author Gender}
\label{sec:persistentgender}
Using name-based inference of first-author gender (resolved for 2,143,536 female and 2,672,411 male first-authored articles, a 71.2\% overall resolution rate), the female share of first authorship rose steadily from 38.9\% (2011) to 46.6\% (2025), a pre-FAIR mean of 40.4\% versus a post-FAIR mean of 44.9\%, never reaching parity with male first authors. Pooled across the whole period, male first authors score slightly higher on average (female mean 4.013 vs male 4.070, gap $-0.057$; Mann-Whitney $p<2.2\times10^{-16}$,
$r=-0.015$, a negligible effect size despite the extreme p-value -- with $n$ in the millions, even a practically meaningless score gap becomes formally ``significant''.

\begin{figure}
\centering
\includegraphics[width=1.0\textwidth]{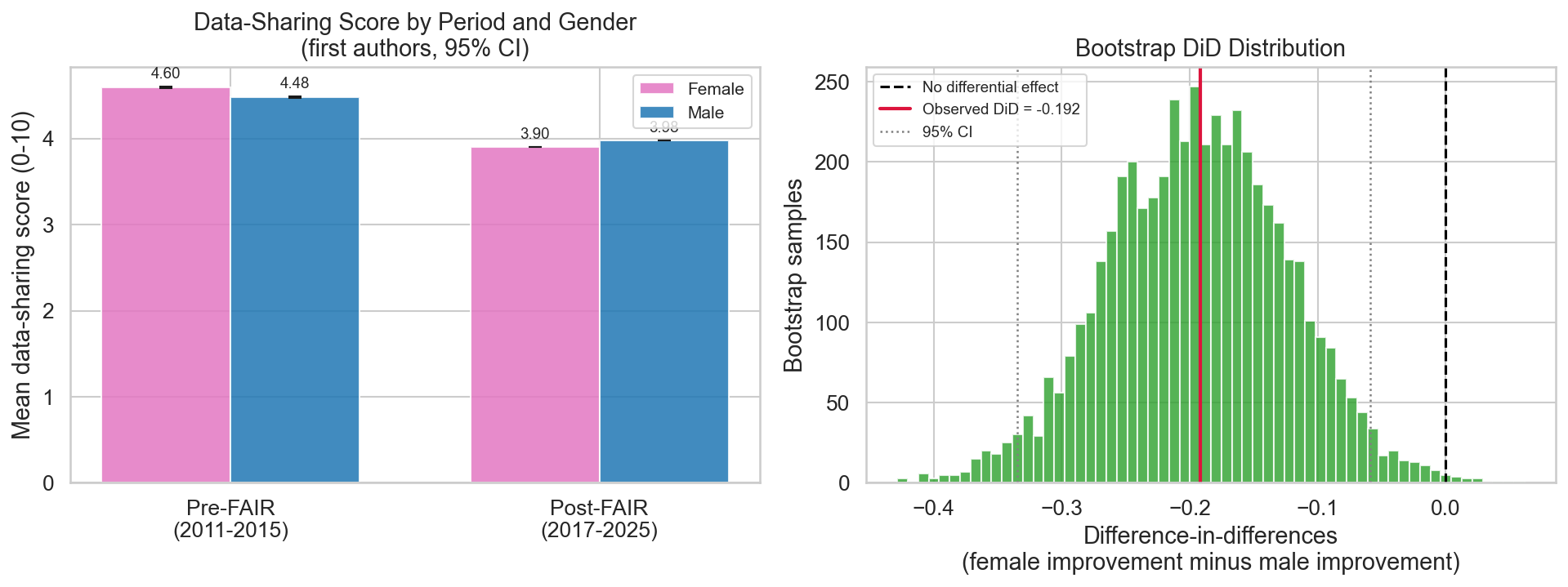}
\caption[]{Data-Sharing Score by Period and Gender (first authors, 95\% CI) and Bootstrap DiD Distribution}
\label{fig:fig_g03_did}
\end{figure}

A formal difference-in-differences analysis shows this gap itself shifted sign around the FAIR Principles' publication (Figure~\ref{fig:fig_g03_did}): the female-minus-male gap was slightly \emph{positive} pre-FAIR ($+0.116$) but turned negative post-FAIR ($-0.076$), with both groups' absolute scores declining (female $-0.693$, male $-0.501$) as part of the corpus-wide decline documented in Section~\ref{sec:thefairscoreparadox}. The DiD estimate is $-0.192$, with a bootstrap 95\% confidence interval ($-0.335$ to $-0.059$, one-sided bootstrap $p=0.0032$) that excludes zero: female first-authored articles' scores declined significantly \emph{more} than male first-authored articles' scores across this shift, even though the two groups started from a near-identical (female-favouring, if anything) baseline rather than an already-diverged one.

The gap does not narrow under international collaboration: the female-minus-male data-sharing-score gap is $-0.049$ for domestic-only first-authored articles (female mean 3.948, $n=$1,806,053; male mean 3.997, $n=$2,237,766) versus $-0.087$ for internationally co-authored ones (female mean 4.361, $n=$337,483; male mean 4.448, $n=$434,645): the gap widens slightly rather than narrowing. Looking beyond first authorship, the same gap is present, and if anything slightly larger, across every author position: pooled over 13,391,265 female- and 20,094,003 male-classified author-article pairs of any position, the gap is $-0.100$ (female mean 4.112, male mean 4.212; Mann-Whitney rank-biserial $r=-0.024$, $p<2.2\times10^{-16}$), compared with $-0.057$ for first authors alone. 
The first-author-specific framing used throughout this subsection therefore does not inflate the gap relative to authorship as a whole; if anything, it slightly understates it.

The gap also varies by discipline: microbiology, plant biology, and clinical research are the disciplines where female first authors outscore male ones (microbiology by the largest margin, $+0.058$; plant biology $+0.045$; clinical $+0.008$), while bioinformatics shows the largest male-favouring gap ($-0.233$). Disciplines with the highest female representation (public health $-0.082$, immunology $-0.003$) still show a negative or near-zero score gap, indicating higher representation alone does not close the gap.

An audit of the underlying name-based inference procedure shows resolution confidence remains high among classified names (mean 0.97) even as the overall resolved share declined gradually over the study period, absorbed mainly by a growing ``ambiguous'' category. Because this decline in resolution is not uniform across countries (name databases skew toward Western naming conventions, so authors from regions under-represented in those databases are disproportionately pushed into ``ambiguous''/``unknown'' rather than resolved, the same coverage bias by nationality reported for SSA-based name matching by \cite{west2013role}), 
a shift in the corpus's geographic composition over 2011--2025 could itself produce an apparent change in the resolved gender distribution or trend, independent of any real change in authorship. This is a confound we have not separated out from the reported gender trend, and it should be read as a caveat on the temporal claims above rather than something the present analysis rules out.

\subsection{Reproducibility Signals and Their Relation to Data-Sharing Score}
\label{sec:limits}
\begin{figure}
\centering
\includegraphics[width=1.0\textwidth]{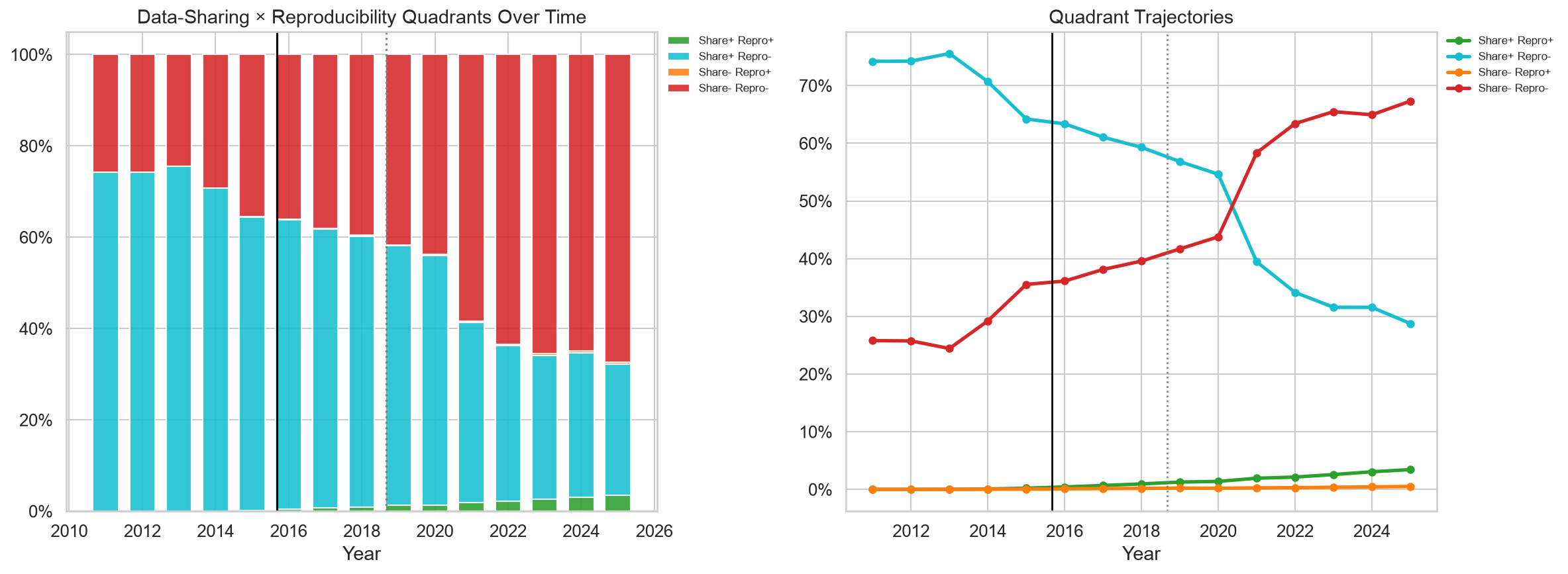}
\caption[]{Reproducibility--FAIR Gap Analysis: Four-Quadrant Populations Over Time. Left: the four Share/Repro quadrants as a \% of articles per year (stacked). Right: the same four quadrants' trajectories plotted as separate lines. Share+/Repro- dominates throughout and grows after 2020 as the Share-/Repro- share falls; the Share+/Repro+ and Share-/Repro+ quadrants both remain small minorities across the whole period. 
}
\label{fig:fig_repro_07_quadrants}
\end{figure}

FAIR compliance and computational reproducibility are conceptually distinct and only weakly coupled empirically, echoing \cite{codebeyondfair2026}'s broader argument that FAIR data principles alone do not guarantee reusable research software, which needs its own, complementary roadmap.
Across six reproducibility signal layers (code sharing, workflow managers, environment specification, version pinning, interactive notebooks, pre-registration), code sharing is by far the most common (1.64\% of the corpus); only 2.1\% of articles have any of these signals, and a ``high reproducibility'' signature ($\geq$3 layers) is present in just 0.006\%. The Pearson correlation between the composite reproducibility score and data-sharing score is positive but modest (r=0.195, $p<0.001$). 
A quadrant analysis crossing high-data-sharing-score status against any reproducibility signal makes this gap concrete (Figure~\ref{fig:fig_repro_07_quadrants}): the ``gold standard'' Share+/Repro+ group is a small minority (1.8\%, mean score 8.174); Share+/Repro- (open data, no code) is the largest 
minority 
(45.2\%, mean 6.241), while Share-/Repro- (neither) is the single largest group overall (52.7\%, mean 1.972): a majority of the corpus achieves neither high-score data sharing nor any detectable reproducibility signal.

Where code is shared, the data-sharing-score uplift is dramatic: articles with a code-repository signal score 7.397 on average versus 3.961 for articles with no code signal at all (Mann-Whitney rank-biserial r=0.770, among the largest effect sizes in this study),
and notebook sharing concentrates almost entirely in the top data-sharing-score quartile, a marker of an already-high-scoring population rather than a broad-based driver. This decoupling of formal compliance from genuine reusability recurs elsewhere: DAS text mentions an explicit licence in only 0.4\% of articles despite article-level licence metadata being present for essentially the whole corpus (100.0\%), so most of the corpus's apparent R1.1 (licensing) compliance is inherited from journal-wide policies rather than specified in the DAS itself; a composite journal-level data-sharing-culture index combining mean score, named-repository rate, and notebook rate shows wide dispersion even among journals with nominally similar DAS policies; and author-level trajectories
show only 30.7\% of authors publishing both before and after 2016 improved their mean data-sharing score, with high-productivity authors clustering in a narrow, slightly negative band that deepens with productivity (mean delta $-0.504$ for authors with 2 papers down to $-0.648$ for those with 26+), a regression-to-the-mean pattern that underscores how little of the corpus-wide data-sharing narrative is being driven by individual authors visibly improving their own practice over time.

\section{Discussion} 
\label{sec:discussion}

\subsection{Principal Findings}
\label{sec:discussion-findings}

Three results recur across almost every analytical dimension explored in this study (discipline, geography, funder, OA tier, gender, and publisher) and together form this paper's central finding. 

First, formal data-sharing documentation and the underlying \emph{quality} of that documentation moved in opposite directions over 2011--2025: explicit DAS adoption rose more than eightfold, while the mean data-sharing score fell (Section~\ref{sec:thefairscoreparadox}), and JATS4R-compliant machine tagging rose from near zero to roughly a third of DAS-bearing articles while remaining uncorrelated (if anything, inversely correlated) with DAS content quality (Section~\ref{sec:machine}). 
Second, this decline is compositional rather than a uniform retreat: existing journals' own named-repository rates barely moved (median +1.1 percentage points per journal), so the corpus-wide pattern is driven by which new articles and venues entered the corpus after 2016, each carrying the restrictive ``available on request'' anti-pattern at a higher rate than the pre-2016 corpus did (Section~\ref{sec:fromsilence}) -- one candidate mechanism behind that growth being the citation advantage repository-linked articles accrue \cite{colavizza2020citation}, as noted in the Introduction.
Third, the decline is pervasive but not absolute: Diamond OA journals and roughly a third of individual journals moved against the corpus-wide trend, showing that a FAIR-supportive trajectory remained achievable throughout the study period for specific communities even as the aggregate picture worsened.

A second cluster of findings concerns where responsibility for these gaps actually sits. JATS4R tagging compliance is essentially entirely at the publisher/production-pipeline level, and 
among the 644 publishers in our corpus with $\geq$50 articles and at least one DAS-bearing article, 414 (64.3\%) sit at an intermediate compliance percentage rather than exactly 0\% or 100\% -- only 117 are fully compliant and 113 fully non-compliant, and 
Table~\ref{tab:jats4r_publisher} gives examples from both ends of the spectrum. Sorting by article volume instead of compliance rate (Table~\ref{tab:jats4r_publisher}, third block) shows this is not a volume effect either: among the ten largest publishers in the corpus, compliance ranges from 0.0\% (MDPI, Cureus Inc.) to 99.3\% (Frontiers Media SA), so the production choice that determines compliance is independent of how many articles a publisher puts through it.
The lowest-scoring repository destinations are disproportionately the plain-\texttt{http://} and legacy \texttt{ftp://} links that depend on third-party infrastructure rather than author effort (Section~\ref{sec:repositorychoice}). This is consistent with the idea that responsibility for reproducibility is not limited to individual authors \cite{codebeyondfair2026}, and extends it: in this corpus, a meaningful share of the machine-readability gap is attributable to publisher XML pipelines and repository infrastructure quality, not to what authors write in their DAS.

Funder and journal mandates show a genuinely double-edged effect rather than a simple success or failure: the ``fully compliant'' share (which presumably represents mostly intended effects) rose steadily throughout the study period, but a much larger share of the corpus was (presumably by way of side effects) pushed down from ``compliant'' into ``partial'' compliance over the same years (Section~\ref{sec:funder}), and no major funder's own pre/post-mandate-year comparison showed a clear, statistically robust improvement once the corpus-wide secular decline is accounted for. 
Mandates appear to be raising the ceiling for a committed minority, while the median article drifts downward, a pattern invisible to any metric that reports only an aggregate ``compliant'' percentage without also tracking its internal composition.

Finally, this study documents two persistent gaps that mandates and recommendations have not closed: the first is a gender gap in data-sharing score that reversed sign (Section~\ref{sec:persistentgender}), going from a small female-favouring difference before the FAIR Principles to a larger male-favouring one after, with female first-authored articles' scores declining significantly more than male first-authored articles' scores across that shift. The second gap is a near-total decoupling of FAIR compliance from computational reproducibility (Section~\ref{sec:limits}), where only 2.1\% of the corpus carries any reproducibility signal at all and the correlation between reproducibility and data-sharing score is positive but modest (r=0.195), consistent with \cite{samuel2026reproscore}'s separate finding that static readiness signals and actual execution-based reproducibility outcomes are only weakly linked in research software more generally. 

\subsection{Implications}
\label{sec:discussion-implications}

For \textbf{authors}, the single most actionable practice this study's data supports is a narrow one: name a specific repository and give it a persistent, resolvable identifier (a DOI or accession number) directly in the DAS, rather than writing only ``available on request'' or pointing to a supplementary file. The difference in outcome is large, not cosmetic: DAS classified as naming a repository predominantly resolve to gold- or silver-tier infrastructure (Section~\ref{sec:repositorychoice}), while the ``available on request'' anti-pattern -- which carries none of that infrastructure benefit -- has itself grown more than 2,250-fold in absolute terms since 2011, faster than the named-repository best practice has grown (Section~\ref{sec:fromsilence}). The same practice extends to code: this study's own prior work on Jupyter notebooks from biomedical publications found that sharing code is not the same as that code remaining executable \cite{samuel2024computational}, and at corpus scale here, only 2.1\% of articles carry any detectable reproducibility signal at all, with code sharing alone the most common of those signals (Section~\ref{sec:limits}); citing a stable, versioned repository identifier for code and workflow artefacts, not only for data, is the same fix applied to a second, parallel gap.
The observed gaps in data sharing~-- particularly those based on discipline, gender and on international collaboration~-- suggest it might be beneficial for researchers to include data sharing aspects more profoundly in their exchanges with colleagues.

For \textbf{publishers}, since JATS4R tagging compliance is determined at the publisher/production-pipeline level, 
this means that the fastest route to closing the machine-readability gap regarding DAS is not individual author education but a comparatively small number of production-pipeline fixes: a single large publisher moving its compliance rate substantially (the extremes in Table~\ref{tab:jats4r_publisher} show this can mean 0\% or 100\%, but most publishers sit somewhere between) would shift the corpus-wide rate more than any plausible author-facing intervention could. The same logic applies to ethics-statement tagging, which lags DAS tagging at every stage (Section~\ref{sec:machine}).

For \textbf{editors and reviewers}, the parameters studied here translate directly into a short, checkable review list that does not require subject-matter expertise: does the DAS name a specific repository and give an accession number or identifier, rather than only asserting that data exists or is available on request (Section~\ref{sec:fromsilence})? Does that identifier actually resolve to a gold- or silver-tier, actively maintained repository rather than a plain-\texttt{http://} or legacy \texttt{ftp://} link (Section~\ref{sec:repositorychoice})?
Is the DAS tagged per JATS4R so it is machine-readable downstream (Section~\ref{sec:machine})? 

Most of the gaps this study documents are visible at review time (with JATS4R compliance being a common exception), yet they persist because the DAS tends to be read less carefully than the rest of the methods sections, not because the underlying problems are hard to see once looked for.

For \textbf{funders and policymakers}, the bifurcation documented in Section~\ref{sec:funder} (fully compliant rising while broadly compliant falls) suggests that mandate design should track more than a single, coarse compliance rate: a policy that looks successful on a binary ``compliant or not'' metric can coexist with the median article's data-sharing quality falling. Tying mandate compliance explicitly to repository tier (gold-tier links already reach 97.0\% mandate compliance versus 26.0\% for bronze-tier hosts, Section~\ref{sec:funder}) or to the four-criterion completeness score used here would give a finer-grained signal than DAS presence alone.

For \textbf{disciplinary societies and other community infrastructure}, the degree of discipline-specific variation documented throughout this study (structural biology's post-2016 decline is the shallowest of any discipline rather than a reversal, Section~\ref{sec:disciplinary}; microbiology and plant biology favour female first authors while bioinformatics shows the largest male-favouring gap, Section~\ref{sec:persistentgender}) means generic, discipline-agnostic FAIR guidance is competing with communities that already have their own vocabularies, repositories, conferences, and professional networks. Scholarly societies and national research-data consortia such as the German National Research Data Infrastructure NFDI \cite{Hartl2021Nationale} are better positioned than any single funder or publisher to make discipline-appropriate repository choice, not just DAS presence, the norm within their own community, using the same channels -- conference guidelines, journal partnerships, discipline-specific training -- already used to disseminate other disciplinary standards.

For \textbf{repository infrastructure}, the F-UJI results (Section~\ref{sec:repositorychoice}) and the gold/silver-tier gap in mandate compliance both point in the same direction: persistent, resolvable identifiers tied to actively maintained, well-described repositories drive most of the measurable FAIR gain, while plain-HTTP and FTP links, and DAS text that says only ``available on request'' without naming a destination at all, drive most of the measurable loss.
Infrastructure investment and guidance (for authors but also editors and reviewers) that steer data toward gold-tier, domain-appropriate repositories, as already happens for much of molecular biology, is likely to move the needle further than guidance aimed only at increasing DAS \emph{presence}, since presence has already risen sharply without a corresponding rise in quality aspects of data sharing.

For \textbf{equity and outreach}, the persistent, and for gender worsening, gaps documented here (by income group, by country, by gender, and by international-collaboration status; Sections~\ref{sec:disciplinary} and~\ref{sec:persistentgender}) complement other observations of social bias around data sharing (e.g.\ \cite{Acciai2023Estimating}) and suggest that FAIR-support efforts framed as purely technical or infrastructural will miss a real part of the problem. The gender gap is worth flagging on its own because it runs counter to an intuitive expectation: international collaboration does not dilute it, and if anything widens it slightly, from $-0.049$ for domestic-only first-authored articles to $-0.087$ for internationally co-authored ones (Section~\ref{sec:persistentgender}), so cross-border co-authorship is not, on its own, a substitute for targeted outreach. 
The disciplinary variation in terms of where female first authors outscore or trail male first authors is specific enough (microbiology and plant biology favour female first authors; bioinformatics shows the largest male-favouring gap) to be directly useful for discipline-specific outreach, for example through the disciplinary societies and consortia named above, rather than only through generic FAIR training.

Diamond OA's trajectory (Section~\ref{sec:diamond}) is this study's clearest positive case study: it is the only access tier whose mean data-sharing score improved, not merely held steady, across the period since the FAIR Principles' publication, and it leads on every secondary quality signal examined. 
Whether this reflects the Diamond OA's community-governed publishing model, selection effects among the research fields and journals that adopt it, or some combination, is a question this paper's descriptive design cannot resolve on its own, but the size of the gap makes Diamond OA's no-APC model a natural subject for follow-up research into \emph{why} it performs differently, including the possibility, raised by \cite{matthias2026diamondfracturestracingjournal},
that journal flipping away from Diamond OA may itself erode part of this advantage over time.

\subsection{Limitations}
\label{sec:discussion-limitations}

\textit{Snapshot retrieval versus a continuously-growing subset.} The PMC-OA subset is not static: articles are added to it after their nominal publication date as embargoes expire, publishers deposit late, or metadata is corrected, so any fixed-date retrieval necessarily undercounts the eventual total for each year, with the undercount largest for the most recently fetched years and smallest for older ones that have had more time to stabilise. We quantified this directly by comparing our per-year article counts against a live NCBI \texttt{esearch} query using the same \texttt{"open access"[filter]} term as the fetch stage. The drift is below 1\% for most years (2012--2014, 2016--2017, 2019--2023), rises to 1.1--2.0\% for 2011, 2015, 2018, and 2024, and reaches 3.2\% 
for 2025, the most recently fetched year, which has had the least time for the OA subset to stabilise. The residual sub-1--3\% drift in these years is retained as an inherent property of any snapshot of a continuously-updated corpus rather than corrected further.

\textit{The data-sharing score is a proxy, not a ground-truth FAIR assessment.} The 0--10 score used throughout most of this paper is derived from DAS category and secondary text signals (named repository, accession number, licence mention, and related features), not from an independent inspection of whether the underlying dataset is actually findable, accessible, interoperable or reusable. 
The F-UJI assessments in Section~\ref{sec:repositorychoice} provide an independent, automated check, but only for the subset of DAS content that resolves to a gold- or silver-tier URL, itself a minority of all DAS-bearing articles; the proxy score and the F-UJI score should be read as complementary, differently biased instruments rather than as the same measurement at different scales.

\textit{Corpus coverage and representativeness.} This study's corpus is based on the PMC Open Access Subset, 
i.e.\ a subset of the scientific literature with a focus on biomedicine.
While we did use some additional services to get complementary information (e.g.\ the Directory of Open Access Journals, as per Section \ref{sec:discipline}), the analyses could potentially have benefited from including additional resources. Some candidate resources for this would be OpenAlex \cite{Forchino2026OpenAlex}, which has information on article processing fees as well as on grant awards, or the Registry of Open Access Repositories Mandatory Archiving Policies (ROARMAP), which has information on institutional and funder policies \cite{swan2015openaccesspolicynumbers}. 

As noted in Section~\ref{sec:corpusgrowth}, the OA subset's own share of PMC has grown from 43.9\% to 91.2\% of PMC's annual output over the study period, so any OA-specific bias toward stronger data-sharing signals should be shrinking rather than remain constant, but the corpus remains PMC-specific (biomedically skewed and predominantly English-language) and does not generalise directly to disciplines, languages or venues outside PMC's scope. 
Separately, the JATS XML tagging~-- despite harmonization upon ingestion into PMC~-- is known to be inconsistent across publishers \cite{mietchen2013inconsistent}, which is precisely the gap Section~\ref{sec:machine} quantifies (and that JATS4R tries to address) rather than a problem this study can fully correct for; any analysis relying on \texttt{sec-type} values (JATS4R compliance, and indirectly the DAS/ethics-statement extraction feeding the rest of this paper) inherits some of that inconsistency as measurement noise rather than true variation in practice.
There have also been reports that the content of the XML versions of articles in PMC may differ slightly
\cite{Serghiou2021Assessment}
or even substantially
\cite{Lawrimore2026Who} from that of the corresponding PDF, including in terms of statements about conflicts of interest \cite{Serghiou2021Assessment} or data sharing \cite{Lawrimore2026Who}.

\textit{Association, not causation.} The funder-mandate (Section~\ref{sec:funder}), discipline (Section~\ref{sec:disciplinary}), and gender (Section~\ref{sec:persistentgender}) analyses are observational comparisons across groups and time, not causal estimates. The difference-in-differences design applied to the gender gap isolates a change in the female-minus-male \emph{gap} relative to the FAIR Principles' timing, a stronger design than a simple before/after comparison, but it cannot rule out a confound that happens to track the same 2016 inflection point and affects the two gender groups differently, for example a shift in the corpus's disciplinary or geographic composition that is itself correlated with gender representation. Similarly, the funder pre/post-mandate-year deltas in Section~\ref{sec:funder} are confounded with the corpus-wide secular decline documented in Section~\ref{sec:thefairscoreparadox}: this study reports that confound explicitly rather than claiming to have isolated a mandate-specific effect.

\textit{Gender inference.} First-author gender is inferred from given names using an automated name-to-gender dictionary, inheriting that method's known biases \cite{santamaria2018comparison,sebo2021performance}: resolution confidence is high among classified names (mean 0.97), but the resolved share declines gradually over the study period, and this decline is not uniform across countries, since name databases skew toward Western naming conventions \cite{west2013role}. 
A shift in the corpus's geographic composition over 2011--2025 could therefore itself produce an apparent change in the resolved gender distribution independent of any real change in authorship; this confound was not separated out and should be read as a caveat on the gender-trend claims in Section~\ref{sec:persistentgender} rather than something this analysis rules out. 

\textit{Reproducibility-signal detection is pattern-based and conservative.} The six reproducibility signal layers used in Section~\ref{sec:limits} (code sharing, workflow managers, environment specification, version pinning, interactive notebooks, pre-registration) are detected from article text and metadata patterns, which is likely to undercount real reproducibility practice not described in a pattern this pipeline recognises, for example code hosted somewhere other than the repositories and mention-patterns searched for. The very low overall rates reported (2.1\% any signal, 0.006\% high reproducibility) should be read as a floor on actual reproducibility, 
not a precise estimate of it. More precisely, in the terminology of \cite{samuel2026reproscore}, these signal layers are static, text-and-metadata markers of reproducibility \emph{readiness} (whether the article mentions code, an environment specification, a workflow manager), not an execution-based measurement of whether that code actually still runs~-- so even this floor should be read as a floor on static readiness specifically, likely still an overestimate of execution-based reproducibility given our own earlier finding that a large share of shared Jupyter notebooks fail to re-execute \cite{samuel2024computational}.

\textit{Assessment of reusability rather than reuse.}
Just as apparent reproducibility readiness does not necessarily translate into execution-based reproducibility, a high score in FAIR assessments of the F-UJI kind does not guarantee actual reuse. The workflow that we used to pipe URLs into F-UJI could, however, be readily modified to search for URLs from DAS in the full text of other articles to identify cases of reuse to analyze.

\subsection{Future Work}
\label{sec:future-work}
We plan to investigate these limitations as part of future work, along with 
exploring some of this study's analytical dimensions (journals, disciplines, funders, demographics, supplementary materials, statement types etc.) in more detail, or zooming in on data contained in our corpus but not assessed in detail in the present study (such as author institutions, software mentioned, or temporal aspects of data sharing policies and practices), or the environmental footprint of such large-scale studies like this one.

Turning this relational dataset into a knowledge graph -- making articles, DAS, repositories, funders, disciplines, and the compliance/score signals reported here explicitly typed, interlinked, and SPARQL-queryable rather than rows in a table -- is natural future work, following the template of our own FAIR knowledge graph built from a smaller computational-reproducibility dataset \cite{samuel2024fair}\cite{samuel2024computational}. Applying the same treatment here would let this paper's specific claims (an article's data-sharing score, its JATS4R tier, its funder's mandate status) be queried and reused directly rather than only read as reported numbers, closing the gap between this study's findings and the FAIR principles it analyses.

\section{Conclusion}
\label{sec:conclusion}
A decade after the FAIR Principles' publication, explicit data availability statements have become the norm rather than the exception across the PubMed Central Open Access corpus, rising from under 2\% of articles in 2011 to 68\% in 2025 (1,969 of 110,347 articles in 2011; 612,057 of 901,213 in 2025, a 311-fold increase in absolute terms).
That growth has not translated into a corresponding rise in data-sharing quality: the mean data-sharing score fell over the same period, exactly compliant JATS4R tagging reaches barely a third of DAS-bearing articles even a decade after the recommendation, 
and the restrictive ``available on request'' anti-pattern grew in step with genuine best practice rather than being displaced by it. Ten years on, writing a data availability statement has become routine; making the underlying data genuinely findable, accessible, interoperable, and reusable has not.

This gap is not evenly distributed. 
Diamond OA journals and overall roughly a third of journals improved or held steady against the corpus-wide decline, showing that FAIR-supportive practice remained achievable throughout the study period rather than becoming structurally impossible, even though every discipline and every one of the top-25-by-volume countries declined overall.
Conversely, low-income countries, several high-volume disciplines (chemistry, immunology, pharmacology), and, after the FAIR Principles' publication, female first-authored articles all saw some steep erosion in data-sharing quality, and funder and journal mandates coincided with a small, fully compliant minority pulling further ahead of a larger majority sliding from ``compliant'' into ``partial'' compliance.
The largest effect size documented anywhere in this study is the data-sharing-score uplift associated with code sharing; Diamond OA's advantage over Gold/subscription is large by comparison with most other splits in this study but clearly smaller than the code-sharing effect. Both point the same direction: where practice does change, the size of the effect can be large, so closing these gaps will need the kind of discipline-, publisher-, and demographic-specific attention this paper's decomposition points toward, rather than a single, generic push for more data availability statements.

More broadly, this study's central methodological point, that the superficial signals that monitoring efforts commonly track (whether a DAS exists, whether it is tagged correctly)
diverge from the substantive signals (especially whether the underlying data is actually reusable), is itself a finding future FAIR-monitoring efforts should take seriously; to our knowledge, this is the first corpus-scale, publicly reported measurement of JATS4R compliance specifically, though PMC as well as individual publishers are plausibly tracking this internally. 
A corpus-wide picture built only from simple compliance measures, without a quality-aware score or an infrastructure-aware check such as F-UJI alongside it, might have reported this decade as one of straightforward progress; the data-sharing score and F-UJI assessments used here show a more complicated reality underneath. 
We make the full pipeline, database, and analysis notebooks openly available.
so that this gap between formal compliance and genuine FAIRness can continue to be tracked, and ideally closed, as the PMC Open Access corpus keeps growing.

\backmatter

\bmhead{Supplementary information}
Not applicable

\bmhead{Acknowledgements}

Our study builds on work by the communities behind Open Access, FAIR, PMC, JATS/JATS4R, Jupyter and related initiatives, for which we would like to express our gratitude. 

D.M. dedicates this work to the memory of Phil Bourne \cite{Prlic2026Phil} and well over a decade of joint activities around the FAIR Principles, data sharing and open scholarship.

\section*{Declarations}

\bmhead{Funding}
This work was supported in part by the German Research Foundation (DFG) through the following projects:
Jupyter4NFDI (\href{https://gepris.dfg.de/gepris/projekt/521453681}{DFG 521453681}  \cite{hagemeier2025Jupyter4NFDI}), 
find.software (\href{https://gepris.dfg.de/gepris/projekt/567156310}{DFG 567156310} \cite{gey2025find.software}), MaRDI (\href{https://gepris.dfg.de/gepris/projekt/460135501}{DFG 460135501} \cite{the_mardi_consortium_2022_6552436} as well as
SeDOA (\href{https://gepris.dfg.de/gepris/projekt/556323977}{DFG 556323977}  \cite{stacker2025SeDOA}) and HYP*MOL (\href{https://gepris.dfg.de/gepris/projekt/514664767}{DFG 514664767}). 
\bmhead{Competing interests}
The authors have no competing interests to declare that are relevant to the content of this article.

\bmhead{Ethics approval and consent to participate}
Not applicable

\bmhead{Consent for publication}
Not applicable

\bmhead{Data availability} 
This study draws on and archives two distinct artefacts, each with its own DOI and licence, reflecting their different natures (our own derived database; our own code):

\begin{itemize}
\item \textbf{Processed database.} The SQLite database (\texttt{pmc\_fair.db}) that backs every number, table, and figure reported in this manuscript -- all extracted data availability statements, ethics statements, classifications, scores, and enrichment variables derived by our pipeline -- is archived separately at Zenodo, \href{https://doi.org/10.5281/zenodo.24209894}{DOI: 10.5281/zenodo.24209894} \cite{samuel2026zenododb}, 
released under the \textbf{CC0 1.0 Universal} Public Domain Dedication.
\end{itemize}

\bmhead{Code availability}
The full data-ingestion pipeline, all analysis notebooks, and the scripts used to generate every figure and tabular export in this manuscript are available at \url{https://github.com/myVSR/fair-data-sharing-evolution}, 
released under the \textbf{Apache License, Version 2.0} and archived at Zenodo \citep{samuel2026zenodocode}.

\bmhead{Materials availability}
Not applicable

\bmhead{Author contribution}
S.S.: Conceptualization, Methodology, Software, Data Curation, Formal Analysis, Investigation, Visualization, Writing -- Original Draft, Writing -- Review \& Editing. \\
D.M.: Conceptualization, Methodology, Formal Analysis, Validation, Writing -- Review \& Editing. Both authors read and approved the final manuscript.

\bibliography{main}%

\end{document}